\documentclass[%
 aip,
 amsmath,amssymb,
 reprint,%
]{revtex4-1}

\usepackage{graphicx}
\usepackage{dcolumn}
\usepackage{bm}

\usepackage{braket}
\usepackage{qcircuit}

\usepackage[utf8]{inputenc}
\usepackage[T1]{fontenc}
\usepackage{mathptmx}
\usepackage{etoolbox}

\makeatletter
\def\@email#1#2{%
 \endgroup
 \patchcmd{\titleblock@produce}
  {\frontmatter@RRAPformat}
  {\frontmatter@RRAPformat{\produce@RRAP{*#1\href{mailto:#2}{#2}}}\frontmatter@RRAPformat}
  {}{}
}%
\makeatother
\begin{document}

\preprint{AIP/123-QED}

\title{Two-qubit quantum photonic processor manufactured by femtosecond laser writing}

\author{N.N. Skryabin}
 \email{Nikolay.Skryabin@phystech.edu}
\author{I.V. Kondratyev}
\author{I.V. Dyakonov}
\affiliation{%
 Quantum Technology Centre and Faculty of Physics, M.V. Lomonosov Moscow State University, 1 Leninskie Gory Street, Moscow 119991, Russian Federation
}%
\author{O.V. Borzenkova}
\affiliation{
Russian Quantum Center, Russia, Moscow, 121205, Bol'shoy bul'var 30 building 1
}%
\author{S.P. Kulik}
\affiliation{%
 Quantum Technology Centre and Faculty of Physics, M.V. Lomonosov Moscow State University, 1 Leninskie Gory Street, Moscow 119991, Russian Federation
}%
\affiliation{%
Laboratory of quantum engineering of light, South Ural State University (SUSU), Russia, Chelyabinsk, 454080, Prospekt Lenina 76
}%
\author{S.S. Straupe}
\affiliation{%
 Quantum Technology Centre and Faculty of Physics, M.V. Lomonosov Moscow State University, 1 Leninskie Gory Street, Moscow 119991, Russian Federation
}
\affiliation{
Russian Quantum Center, Russia, Moscow, 121205, Bol'shoy bul'var 30 building 1
}%

\date{\today}

\begin{abstract}
We present an experimental implementation of a two-qubit photonic quantum processor fabricated using femtosecond laser writing technology. We employ femtosecond laser writing to create a low-loss reconfigurable photonic chip implementing precise single-qubit and two-qubit operations. The performance of single-qubit and two-qubit gates is characterized by full process tomography. An exemplary application of the processor to determining the ground state energy of an $H_{2}$ molecule using the variational quantum eigensolver algorithm is demonstrated. Our results highlight the potential of femtosecond laser writing technology to deliver high quality small-scale quantum photonic processors.
\end{abstract}

\maketitle

\section{\label{sec:level1}Introduction}

The development of technological platforms for quantum computing is actively gaining momentum due to rapidly evolving commercial potential of quantum computing \cite{MacQuarrie2020, GYONGYOSI201951}. Recent quantum advantage demonstration using gaussian boson sampling \cite{Zhong1460} has shown prospect for scaling photonic systems beyond the reach of classical simulation. Commercial companies, like PsiQuantum and Xanadu, \cite{Arrazola2021} lead the research towards a fully scalable fault-tolerant photonic quantum computer. Their approach heavily relies on photonic integration technology. To date, a wide range of integrated photonic platforms are adopted for quantum experiments including silica-on-silicon, silicon-on-insulator, silicon nitride, III-V semiconductors, diamond and diamond-on-insulator, lithium niobate and lithium niobate-on-insulator, femtosecond laser writing (FLW) in silicate glasses and others \cite{Meany2015, Bogdanov17, Flamini2018, Wang2020, Elshaari2020, Adcock2021, Saravi2021}. 

The FLW technology is based on a local and permanent change in the refractive index of transparent dielectric materials under the action of tightly focused ultrashort laser pulses and allows to prototype various kinds of three-dimensional integrated elements and devices in a single step \cite{Tan2021}. The application of the FLW in quantum photonics has begun with the observation of the HOM interference in a directional coupler \cite{Marshall2009}. The same kind of structure enabled on-chip polarization-entangled state measurement \cite{Sansoni2010}. This was followed by several works focused on the implementation of quantum gates for polarization qubits based on polarization beam splitters and waveplates fabricated on a chip\cite{Crespi2011, Corrielli2014, Heilmann2014}. The next milestone were the first boson sampling experiments \cite{Crespi2013, Tillmann2013} based on the FLW passive multichannel interferometers. Further works demonstrated generation of higher-order W-states \cite{Grafe2014}, preparation and measurement of hyper-entangled and cluster states \cite {Ciampini2016}, heralded two-qubit CZ \cite{Meany2016} and CNOT \cite{Zeuner2018} gates for polarization qubits, and a CNOT gate for dual-rail qubits \cite{Zhang2019, Li2020}. The chips used in these works lacked reconfigurability which is essential for linear-optical quantum computing. The thermooptic modulation of the FLW circuits was first explored in a most simple two-mode Mach-Zehnder interferometer \cite{Flamini2015, Chaboyer:17}, and was extended to multiport devices \cite{Crespi2017, Dyakonov2018} later. Recent works target the efficiency of thermooptical phase shifters \cite {Ceccarelli:19, Ceccarelli2020} and scalability of the FLW-fabricated reconfigurable circuits \cite{Hoch2021, Valery2022}.

In this work we fabricate and characterize a programmable two-qubit photonic processor using the FLW technology. Programmable gate operations were only demonstrated with integrated circuits fabricated using silicon photonics technologies \cite{Shadbolt2012, SantagatiNov2017}. Several groups achieved two-qubit gate fidelity reaching 90\% and above \cite{Qiang2018, Feng2022}.  We create a two-qubit processor endowed with programmable single-qubit gates and a fixed two-qubit CNOT gate. The performance of each single-qubit gate and the two-qubit is carefully studied. As an exemplary application we implement the variational quantum eigensolver (VQE) algorithm searching for the ground state energy of the $H_{2}$ molecule.

\section{Methods}
\subsection{\label{sec:level2}Device design and fabrication}

\begin{figure*}[t]
\centering
\includegraphics[width=2.0\columnwidth]{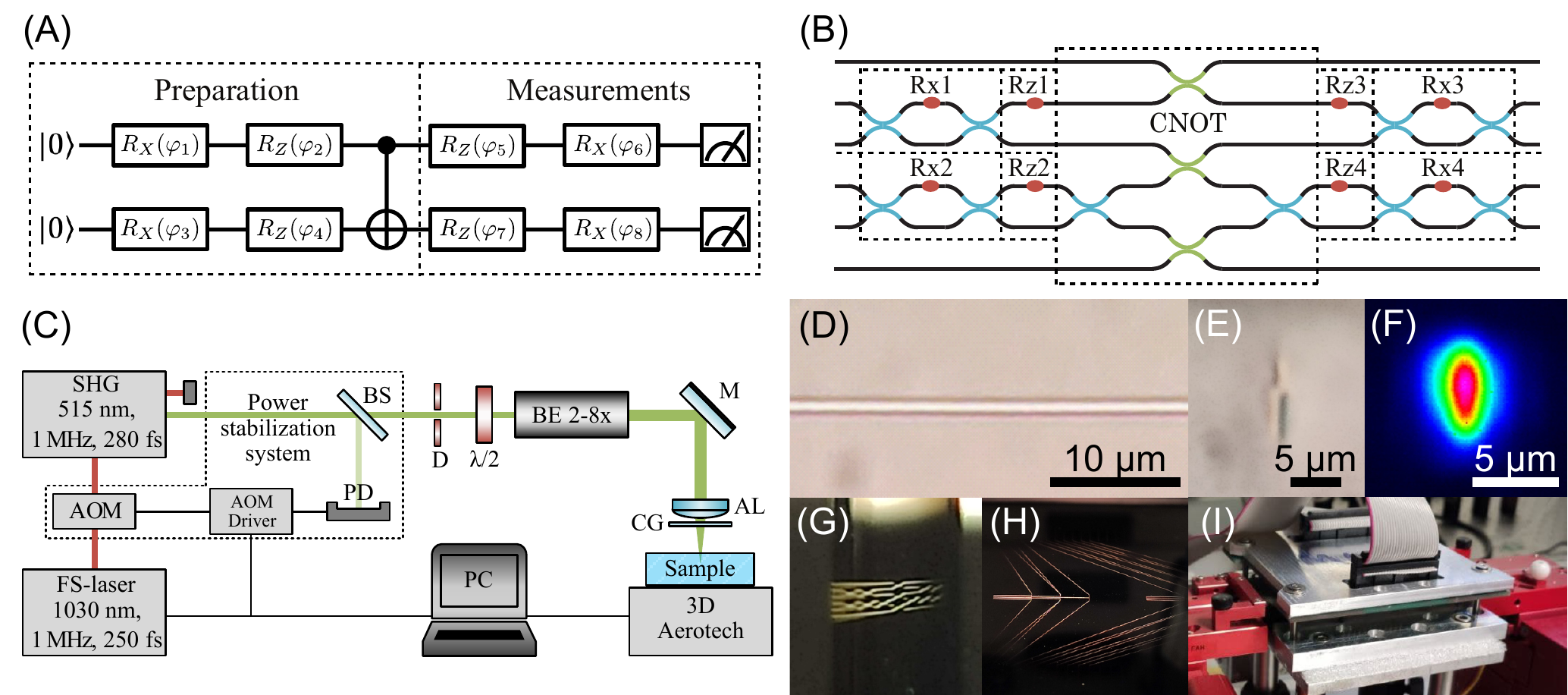}
\caption{\label{fig::U2q} (\textbf{a}) The quantum circuit of the photonic processor. (\textbf{b}) The optical circuit of the processor (blue and green directional couplers denoting splitting ratio 50:50 and 33:67, respectively, red ovals represent phase shifters). (\textbf{c}) The experimental setup scheme for femtosecond laser writing. SHG--second harmonic generator, AOM--acousto-optic modulator, $\lambda /2$--half-waveplate, BS--beamsplitter, PD--photodiode, BE--beam expander, M--mirror, AL--aspheric lens ($NA = 0.55$), CG--cover glass, 3D Aerotech -- automated three-axis positioning system. (\textbf{d}) The microscope images of the top and (\textbf{e}) cross-section views of the waveguides. (\textbf{f}) TE mode field profile for the wavelength of 810 nm. (\textbf{g}) The end face photograph of the fabricated waveguide structure inside fused silica sample. (\textbf{h}) The photograph of engraved structures in the nichrome (NiCr) layer deposited onto the surface. (\textbf{i}) The photograph of finished assembly of the processor.}
\end{figure*}

The photonic processor includes two logically distinct parts, Fig.~\ref{fig::U2q}a illustrates the quantum circuit of the processor. The first part includes two single-qubit gates and a two-qubit CNOT gate and serves as a quantum state preparation circuit. The second one implements the single-qubit measurements. We base the design of the optical circuit on the dual-rail encoding of the qubits (see Fig.~\ref{fig::U2q}b). Each single-qubit operation is a sequence of a programmable $R_{x}(\varphi)$ and $R_{z}(\varphi)$ gates. The $R_{x}(\varphi)$ gate is a Mach-Zehnder interferometer which consist of two balanced directional couplers and a phaseshifter which controls the phase $\varphi$. The $R_{z}$ gate is simply a phaseshifter, which sets the phase shift value $\varphi$. The non-deterministic CNOT gate with a success probability 1/9 is designed according to \cite{Ralph2002}. The probabilistic nature of a two-photon CNOT gate forces the use of postselection to rule out the outcomes which are not contained in the logical Fock subspace of the dual-rail encoded qubits.

The waveguide structure of the photonic chip was fabricated by the FLW method in a 100~mm~$\times$~50~mm~~$\times$~5~mm fused silica sample (JGS1 glass material). The scheme of the writing setup is illustrated in Fig.~\ref{fig::U2q}c. Femtosecond laser pulses from a second harmonic of an ytterbium fiber laser (Avesta Antaus, wavelength 1030/515 nm, pulse duration 250/280 fs, pulse energy 400/80 nJ, repetition rate 1 MHz) were focused using an aspheric lens (NA = 0.55) at 15 $\mu$m depth below the surface through a 170 µm cover glass used for partial compensation of spherical aberrations. The variable beam expander increased the laser beam in order to completely fill the entrance aperture of the focusing lens. The sample was translated relatively to the focal volume with a high-precision air-bearing stage (AeroTech FiberGlide3D) at the 0.2 mm/s velocity. The polarization of the laser beam is set parallel to the writing direction by a half-waveplate $\lambda/2$. The microscope images of the top and cross-section views of the inscribed waveguides are presented in Fig.~\ref{fig::U2q}d,e. The waveguides support a single guided mode with mode field diameters 5~$\mu$m~$\times$~8~$\mu$m (see Fig.~\ref{fig::U2q}e) and exhibit 0.5~dB/cm propagation loss at 810~nm wavelength for a TE polarized mode. Directional couplers consist of two waveguides with two s-bends, which bring cores together to the distances of 7.62~$\mu$m and 6.86~$\mu$m corresponding to devices with 50:50 and 33:67 power splitting ratios, respectively. The s-bends are the combination of two circular arcs with radius 80~mm and introduce a slight $\textless$ 0.1 dB/cm additional bending loss. The distance between the waveguides at the input/output facet and inside the circuit was set to match the fiber array pitch of 250~$\mu$m. After the waveguide structure had been inscribed, the input/output facets of the sample were optically polished  The end face photograph of the fabricated waveguide structure inside fused silica sample is shown in Fig.~\ref{fig::U2q}g.  

The reconfigurability of the quantum processor was introduced by employing a thermo-optical effect. We deposited the nichrome (NiCr) layer with thickness of about 0.1~$\mu$m onto the surface of the photonic chip by the magnetron sputtering method. Next we used the same FLW setup to locate alignment markers and engrave the heaters that are 30 $\mu$m wide and 3~mm long, the contact pads, and the connecting wires in precisely designed locations (see Fig.~\ref{fig::U2q}h). The values of electrical resistances of the fabricated heaters varied from 0.9 to 1.2 k$\Omega$ depending on the thickness of the NiCr layer. The electrical connection was established through the printed circuit board (PCB) with spring-loaded contacts. The photograph of the finished assembly of the photonic processor is shown in Fig.~\ref{fig::U2q}i. The home-built 16-channel 12-bit digital constant current source supplied the power to drive the heaters. The photonic chip rested on the aluminum stage which was actively stabilized and kept a constant $20^{\circ}$~C temperature of the bottom surface of the chip.

\subsection{Experimental setup}

\begin{figure*}[t]
\centering
\includegraphics[width=2\columnwidth]{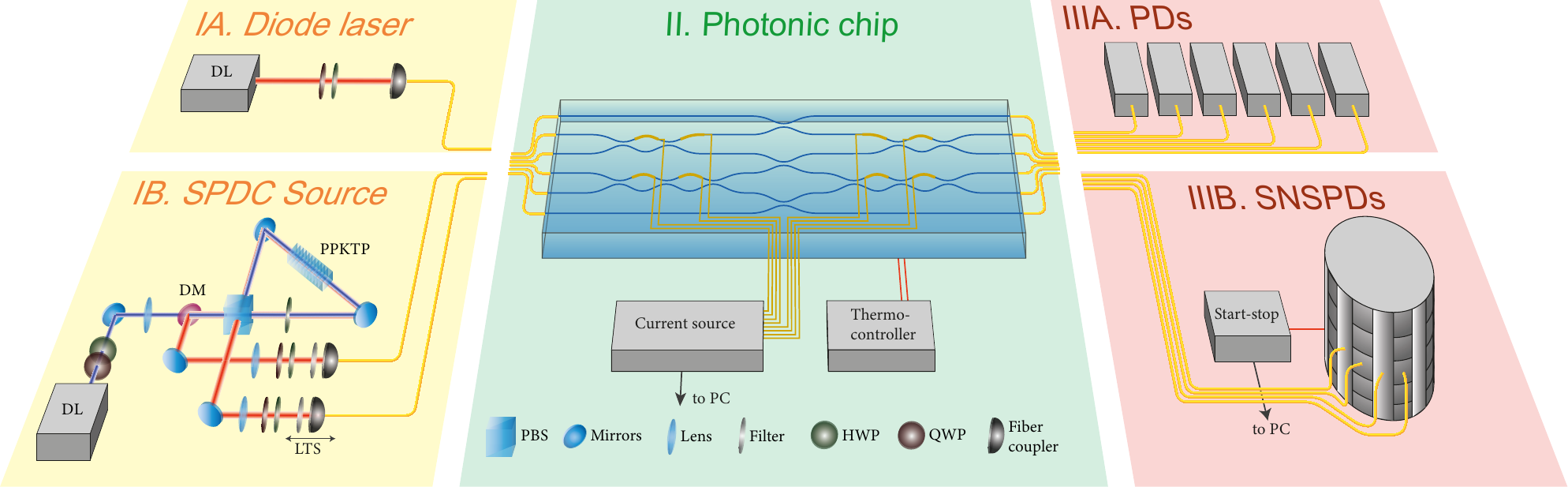}
\caption{\label{fig::ExpSetup} The scheme of classical (IA-II-IIIA) and quantum (IB-II-IIIB) experimental setup. DL--diode laser, SPDC--spontaneous parametric down-conversion, DM--dichroic mirror, PPKTP--periodically poled potassium titanyl phosphate, LTS--linear translation system, PBS--polarization beam splitter, HWP--half-waveplate, QWP--quarter-waveplate, PDs--photodetectors, SNSPDs--superconducting nanowire single-photon detectors, PC--personal computer.}
\end{figure*}

The optical circuit of the processor was first tested using a CW diode laser at 808~nm (see Fig.~\ref{fig::ExpSetup}, set IA-II-IIIA). The V-groove fiber arrays were used to launch the laser radiation into the chip and to collect light at the output facet. The pair of quarter- and half-waveplates (QWP and HWP) tuned the input polarization to vertical before launching. Six home-built PIN-diode-based photodetectors registered the output power at each port of the chip.

Experiments in the quantum regime (see Fig.~\ref{fig::ExpSetup}, set IB-II-IIIB) were carried out using the spontaneous parametric down-conversion (SPDC) source of photon pairs and commercially available superconducting nanowire single-photon detectors (SNSPDs, Scontel). The core of the single-photon source was a 30-mm long periodically poled potassium titanyl phosphate (PPKTP) type-II nonlinear crystal placed in the Sagnac-type optical scheme~\cite{zeilinger2007source}. The source was pumped by the CW wavelength-stabilized diode laser (Ondax SureLock, 6~mW output power, 405~nm wavelength, 1~MHz linewidth) and generated spectrally degenerate signal and idler photons at 810~nm wavelength with orthogonal polarizations. The polarization of the pump beam was set to generate factorized polarization state. The photons were split on the polarizing beamsplitter (PBS) and guided through different optical paths of the scheme. Each path was coupled to a single-mode fiber using an aspheric lens. The source delivered 600 kHz single-photon count rate and 80 kHz coincidence count rate. The motorized quarter- and half-waveplates (QWP and HWP) placed in front of the fiber couplers compensated the polarization rotation happening in the single-mode fibers connected to the input of the chip. The output fibers were connected to the SNSPDs. The 4-channel start-stop electronic circuit registered coincidence counts within a 4~ns window.   Lastly, we synchronized the arrival of both photons to the chip using the translation stage (LTS) mounted under one of the fiber couplers. Before running the experiment  we benchmarked the source by measuring the Hong–Ou–Mandel (HOM) interference visibility $V_{HOM}$ inside the chip. We obtained a value $V_{HOM}=95.7$~\% which signals about the good quality of the source (see Appendix~\ref{app:photons_indist_tune}). This value is also used to compare results to the simulation of the experiment which helps to establish main sources of the photonic processor defects (see Appendix~\ref{app:chip_simul} for simulation details).
 
\section{Experiment and results}

\subsection{Classical characterization of the device}

\begin{figure}[h]
\centering
\includegraphics[width=1.0\columnwidth]{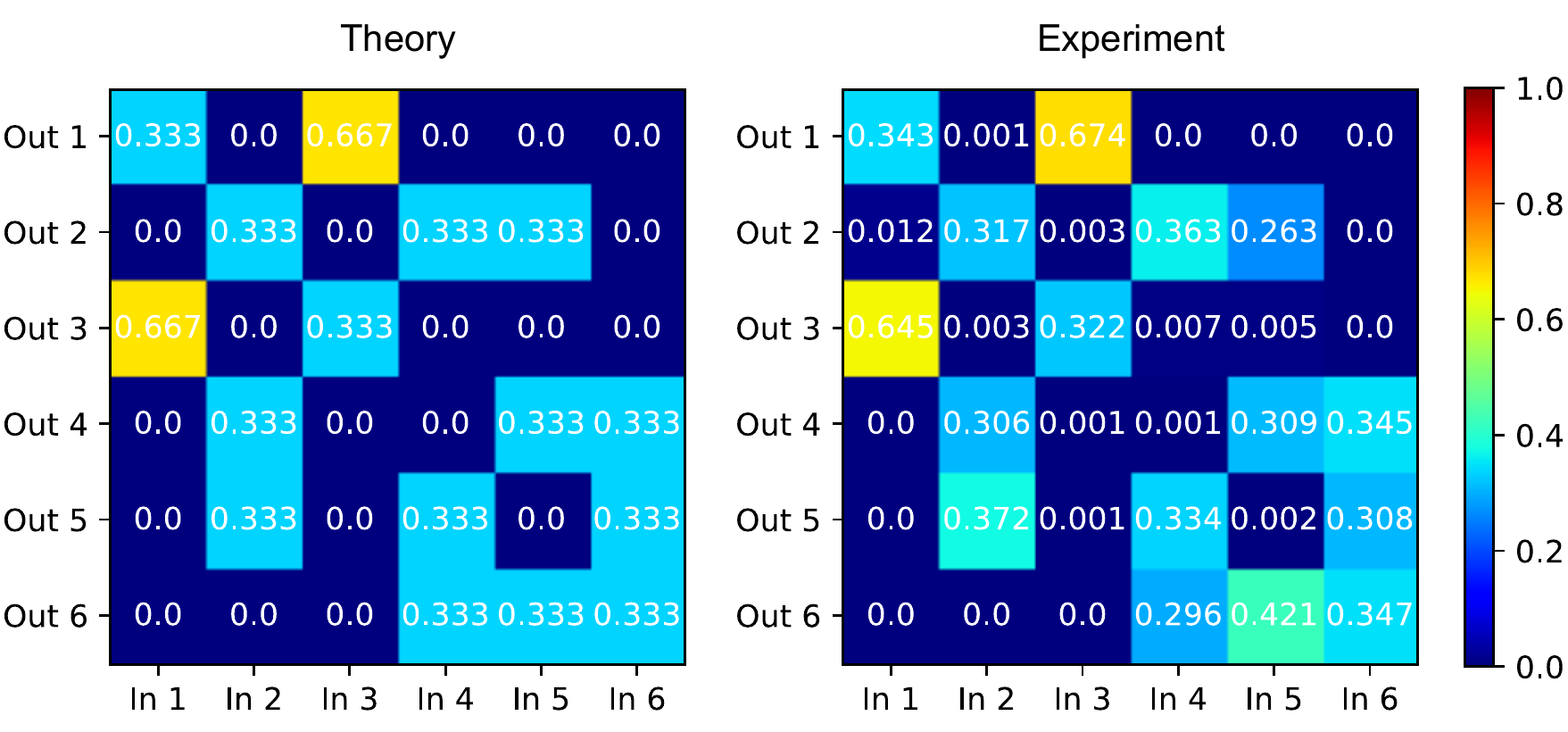}
\caption{\label{fig::iSST} Theoretical ideal (left) and experimentally obtained (right) transformation matrices. Fidelity with ideal 99.18 \% }
\end{figure}

At the first step the passive optical circuit was characterized with coherent 808~nm laser light. The laser radiation was injected consequently into all input modes one by one and the output power at all of the six modes of the circuit was measured. Afterwards the measured data were fed to the Sinkhorn-Knopp algorithm which yielded the matrix of the square moduli $|U_{ij}|^{2}$ of the passive circuit unitary transformation elements. The experimentally obtained $|U_{ij}|^{2}$ matrix and the theoretical ideal counterpart are shown in Fig.~\ref{fig::iSST}. The fidelity between these two matrices is $F=99.18$~\%, which was computed according to the expression
\begin{equation}
\label{fidelity_formula}
F(U_{e}, U_{t}) = \frac{|Tr(U_{e}^{\dagger}U_{t})|^2}{Tr(U_{e}^{\dagger}U_{e})Tr(U_{t}^{\dagger}U_{t})}.
\end{equation}

This result clearly indicates that the directional couplers comprising the circuit satisfy the designed ratios with a small error. We cannot directly extract the exact values of the directional couplers' ratios but the overall performance of the passive circuit falls within the expected quality level. This measurement also revealed that the propagation loss is homogeneous inside the circuit (which means there are no internal defects inside the circuit). The total loss across all paths lies in the $6.8 - 6.9$~dB range. 

\subsection{Phaseshifter calibration and thermal cross-talk compensation}

The photonic processor is equipped with eight thermooptical phaseshifters which control single-qubit operations (see Fig.~\ref{fig::ExpSetup}a). The precise operation of the processor requires calibrating the phase $\phi(I)$ dependence on the applied current for each modulator. The calibration of the selected phaseshifter starts with injection of the laser light into the appropriate input port and continues with the power measurement on each of the output ports while sweeping the current running through the selected heater. Ideally the normalized output power $P$ from a single output mode should depend on the applied current value $I$ as:

\begin{equation}
\label{eqn:calib_fit}
P(I) = a + b \cos{(\phi_0 + \alpha I^2)},
\end{equation}
where constants $a,b$ are defined by the optical circuit configuration and $\phi_0$ is a constant phaseshift between the optical paths. The $\phi_{0}$ phaseshifts in our circuit were designed to be equal to $0$, however the imperfections of the fabrication procedure or the sample material induce parasitic non-zero phaseshifts. Furthermore, the phaseshifters exhibit large thermal cross-talk due to the relatively small distance between them compared to their width. Hence, the calibration procedure has to compensate for the cross-talk effect. If the parasitic thermal cross-talk is present, then the phase on the $i$-th phaseshifter will be described by the equation:

\begin{equation}
\label{eqn:phase_crosstalk}
\phi_i = \phi_{0i} + \sum\limits_{j}\alpha_{ij} I_{j}^2,
\end{equation}
where $I_j$ is the applied electrical current on the j-th heater. The $\alpha_{ij}$ matrix describes the effect of the $j$-th heater on the $i$-th phaseshifter. In our circuit we arranged the heaters in pairs one above the other, so the thermal cross-talk effect between them can be compensated by the heaters themselves. The required currents $I = \{i_1,...,i_n \}$ are the solution of the system of equations:
\begin{equation}\label{eqn:cross-talk_phase_calculation}
    I = \sqrt{ \alpha^{-1} \times (\Phi - \Phi_0 )},
\end{equation}
where $\Phi$ is a vector of the required phaseshifts on each heater, $\Phi_0$ is an initial phase shifts vector and $\alpha$ is the heater effect coefficient matrix. More details on the calibration procedure can be found in Appendix~\ref{app:calibration}. 

\subsection{Characterization of the single-qubit gates}

Linear-optical quantum computing platform offers a peculiar property of the information carrier --- the qubit is flying --- which precludes from benchmarking the gate performance using standard algorithms \cite{Emerson2011, Nielsen2021}. The optical qubit propagates through a series of gates comprising the processor and it is costly to introduce additional optical elements which can isolate each single- or two-qubit gate. Here we implement a simple operational characterization method which allows us to estimate the quality of a single qubit gate inside an optical processor with a static gate layout. 

Each of the single-qubit gates $R_{x}(\varphi)$ and $R_{z}(\varphi)$ is either a combination of directional couplers and a phaseshifter or simply a phaseshifter alone. We use the knowledge about the components of each gate --- the directional coupler splitting ratios and the phaseshifter calibration --- to infer the best possible fidelity which is reachable with our experimental equipment. In our setup the imperfections of the optical implementation of the single-qubit operations in dual-rail encoding stem from deviations of directional coupler splitting ratios and discreteness of electrical current values. We would like to note that we do not account for any additional source of errors such as electrical noise of the current source or any other effects which can affect the single-qubit gate performance. 

The calibration procedure of the $R_{x}$ gate yields the splitting ratios $r_{1},r_{2}$ of both directional couplers and the calibration curve $\phi(I)$. This data is enough to build a complete model $\mathcal{U_{e}}$ of each $R_{x}$ gate inside the processor. Next we generate a set of 100 random phases $\varphi_{t}$ which correspond to a set of random $R_x(\varphi_{t})$ gates and compute their theoretical $2\times 2$ unitary matrices $U_{t}$. The $\phi(I)$ curve helps us to establish the closest phases $\theta_{e}$ which can be set in the experiment. The model of the $U_{e}=R_{x}(\varphi_{e})$ gate is used to estimate the fidelity value of the experimentally feasible implementation of the gate $U_{t}$. The fidelity was computed according to the equation \ref{fidelity_formula}.
Similarly, we estimate the performance of each $R_{z}$ gate which depends only on $\phi(I)$.

\begin{figure*}[t]
\centering
\includegraphics[width=2.0\columnwidth]{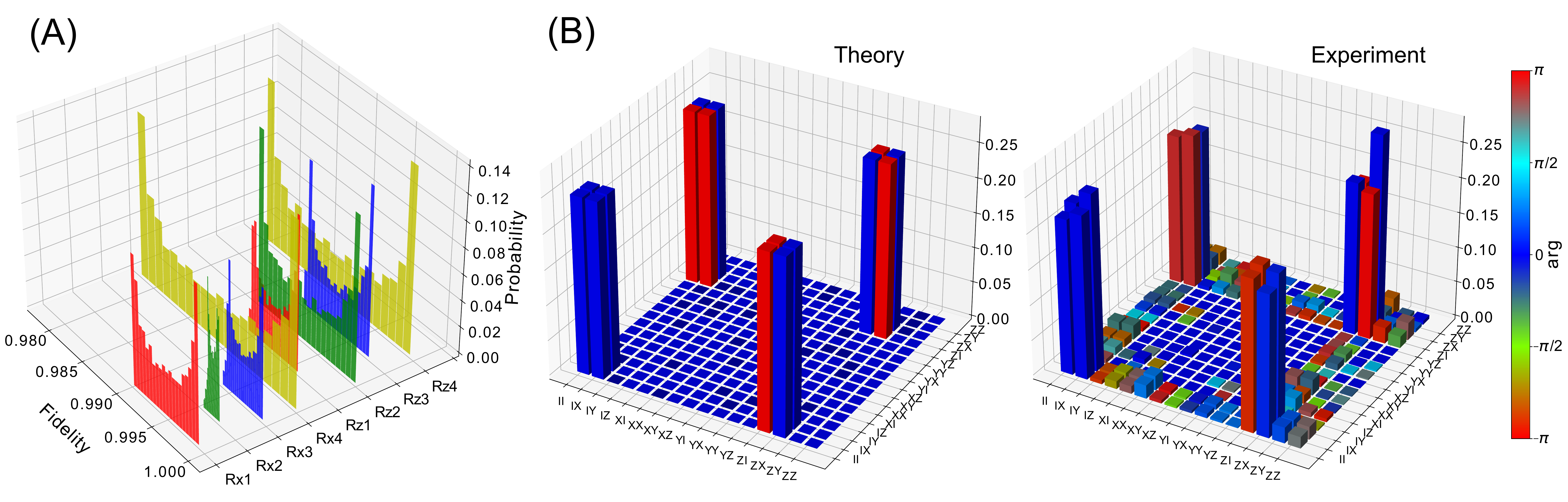}
\caption{\label{fig::gates_fidelity} (a) Histograms of single qubit gates fidelity distributions. Fidelity of all single qubit rotation gates was not less than 97.5 \%. (b) Theoretical ideal two qubit CNOT gate process $\chi$-matrix (left) and obtained experimental process $\chi$-matrix (right). Experimental matrix fidelity with theory is 94.4 \%}
\end{figure*}

We plot the fidelity $F$ histograms for each single-qubit operation in Fig.~\ref{fig::gates_fidelity}a.  The minimal $F$ value encountered across all single-qubit gates was 97.5 \%. Mean fidelities and their standard deviations for each single qubit gate are presented in Table \ref{tab::single_gates_fids}.

\begin{table}[h]
    \centering
    \begin{tabular}{ |c|c|c|c|c|c|c|c|c| } 
     \hline
     Gate & $R_{x1}$ & $R_{z1}$ & $R_{x2}$ & $R_{z2}$ & $R_{x3}$ & $R_{z3}$ & $R_{x4}$ & $R_{z4}$ \\ \hline
     Mean & 0.995 &	0.993 & 0.997 & 0.993 & 0.994 & 0.997 & 0.988 & 0.986  \\ \hline
     Std & 0.003 & 0.003 & 0.001 & 0.005 & 0.003 & 0.002 & 0.007 & 0.008  \\ \hline
    
    \end{tabular}
    \caption{\label{tab::single_gates_fids} Mean fidelities and standard deviations for each single qubit gate.}
\end{table}

\subsection{Characterization of the two-qubit gate}

The layout of the photonic processor allows to run a standard quantum process tomography (QPT) routine in order to establish the quality of the two-qubit CNOT gate \cite{chuang1997prescription}. The QPT reconstructs the $16\times 16$ $\chi$-matrix of the unknown two-qubit quantum process which completely characterizes its' effect. In our case the process under study is the two-qubit CNOT gate. 

In the experiment two indistinguishable single photons were injected into the processor to initialize the logical $\ket{00}$ state. The measurement circuit registered two-fold coincidence configurations corresponding to $\ket{00}$, $\ket{01}$, $\ket{10}$ and $\ket{11}$ states. The single-qubit gates $R_{x1}, R_{z1}, R_{x2}, R_{z2}$ prepared the factorized input states of the two dual-rail qubits. The gates $R_{x3}, R_{z3}, R_{x4}, R_{z4}$ set the measurement basis. We used the two-qubit Pauli decomposition domain, which means the resulting $\chi$-matrix was expressed in the two-qubit Pauli basis (see Appendix~\ref{app:qpt_appdx} for details).   

The following set of states was used as the input to the quantum process:
\begin{gather*}\label{eqn:single_qubit_input_list}
   \ket{0}, \ \ket{1}, \\
   \ket{+} = \frac{\ket{0} + \ket{1}}{\sqrt{2}}, \  
   \ket{-} = \frac{\ket{0} - \ket{1}}{\sqrt{2}}, \\
   \ket{i} = \frac{\ket{0} + i \ket{1}}{\sqrt{2}}, \ 
   \ket{-i} = \frac{\ket{0} - i \ket{1}}{\sqrt{2}}
\end{gather*}

After undergoing the transformation by a quantum process both qubits were projected on the particular state from the similar set $\{\ket{0},\ket{1},\ket{+},\ket{-},\ket{i},\ket{-i}\}$ and photon statistics was measured. We performed 256 linearly independent measurements while the complete measurement set includes at minimum 240 independent projections \cite{chuang1997prescription}.  

The maximum likelihood procedure seeks for an optimal process $\chi$-matrix which fits the obtained measurement results the best. The $\chi$-matrix hypothesis was parameterized with 256 real numbers $\Vec{t}$ according to \cite{altepeter2005photonic}:
\begin{equation}\label{eqn:chi_parametrization}
\chi(\Vec{t}) =  \frac{g(\Vec{t})g(\Vec{t})^{\dagger}}{Tr[ g(\Vec{t})g(\Vec{t})^{\dagger} ]}, 
\end{equation}
where $g(\Vec{t})$ can be an arbitrary complex matrix parametrization, for example, the triangular one \cite{altepeter2005photonic}. The optimization was carried out using the SciPy package. The theoretical and the reconstructed $\chi$-matrices are shown in Fig.~\ref{fig::gates_fidelity}b. To compare the obtained process $\chi$-matrix with its theoretically expected counterpart the matrix fidelity metric (\ref{fidelity_formula}) was used. We achieved the $94.4 \%$ fidelity for the reconstructed $\chi$-matrix of the two qubit CNOT gate in the fabricated processor. We compared this result to the simulation of the CNOT quantum process with different types of possible physical defects affecting the performance of the gate (see Appendix~\ref{app:chip_simul}). The limiting factor turned out to be the indistinguishability of the input photons. It can be clearly inferred by witnessing a characteristic $\chi$-matrix distortion occurring when the input photons are distinguishable.

\subsection{Variational eigenvalue solver}

As an exemplary application demonstrating the capabilities of the fabricated processor, we implemented the variational quantum eigensolver algorithm (VQE) which calculates the value of the bound energy of the hydrogen $H_2$ molecule. A VQE algorithm is used to estimate the expectation value of a Hamiltonian of a quantum system. The algorithm uses a classical optimizer to minimize the measured expectation value $\bra{\Psi} H \ket{\Psi}$. The Hamiltonian $H$ is decomposed into a sum of Pauli matrix tensor products. Firstly, the Hamiltonian of the $H_{2}$ molecule is expressed in terms of the fermionic creation and annihilation operators $\hat{a}$, $\hat{a}^\dagger:$

\begin{equation}\label{eqn:fermionic_hamiltonian}
H_F = \sum_{ij} t_{ij}\hat{a_i}^\dagger \hat{a_j} + \sum_{ijkl} u_{ijkl} \hat{a_i}^\dagger \hat{a_k}^\dagger \hat{a_l} \hat{a_j}.
\end{equation}

The $t_{ij}$ and $u_{ijkl}$ coefficients are computed numerically using an STO-3G basis. Next a mapping from the fermionic operators to the qubit operators is applied. We used the Bravyi-Kitaev transformation to get the Pauli matrix decomposition of the $H_{2}$ hamiltonian:

\begin{eqnarray}\label{eqn:qubit_hamiltonian}
   H_{H_2} = f_0 \mathbb{I} \otimes \mathbb{I} + f_1 \sigma_Z \otimes \sigma_Z + f_2 \sigma_Z \otimes \mathbb{I} + 
   \nonumber \\f_3 \mathbb{I} \otimes \sigma_Z + f_4 \sigma_X \otimes \sigma_X .
\end{eqnarray}

To simplify the form of the Hamiltonian in the qubit representation, the terms with factors less than $10^{-8}$ were discarded. In addition, taking into account only short-range interaction makes it possible to reduce the number of qubits required for an accurate description. In the case of the hydrogen molecule, the two-qubit model is sufficient for precise binding energy estimation. We used the Openfermion software package to get both the $t_{ij}, u_{ijkl}$ and $f_{i}$ coefficients. \\

We used the quantum processor to measure the expectation values of each of the terms in the Hamiltonian (\ref{eqn:qubit_hamiltonian}) and then added them up to obtain the energy expectation value $\langle H \rangle_{e}$. The single qubit gates $R_{x1}, R_{z1}, R_{x2}, R_{z2}$ and the two-qubit CNOT gate prepared the two-qubit probe state $\ket{\Psi}$, which is employed for expectation value $\langle H \rangle_{e}$ estimation.

\begin{figure}[h]
\centering
\includegraphics[width=1.0\columnwidth]{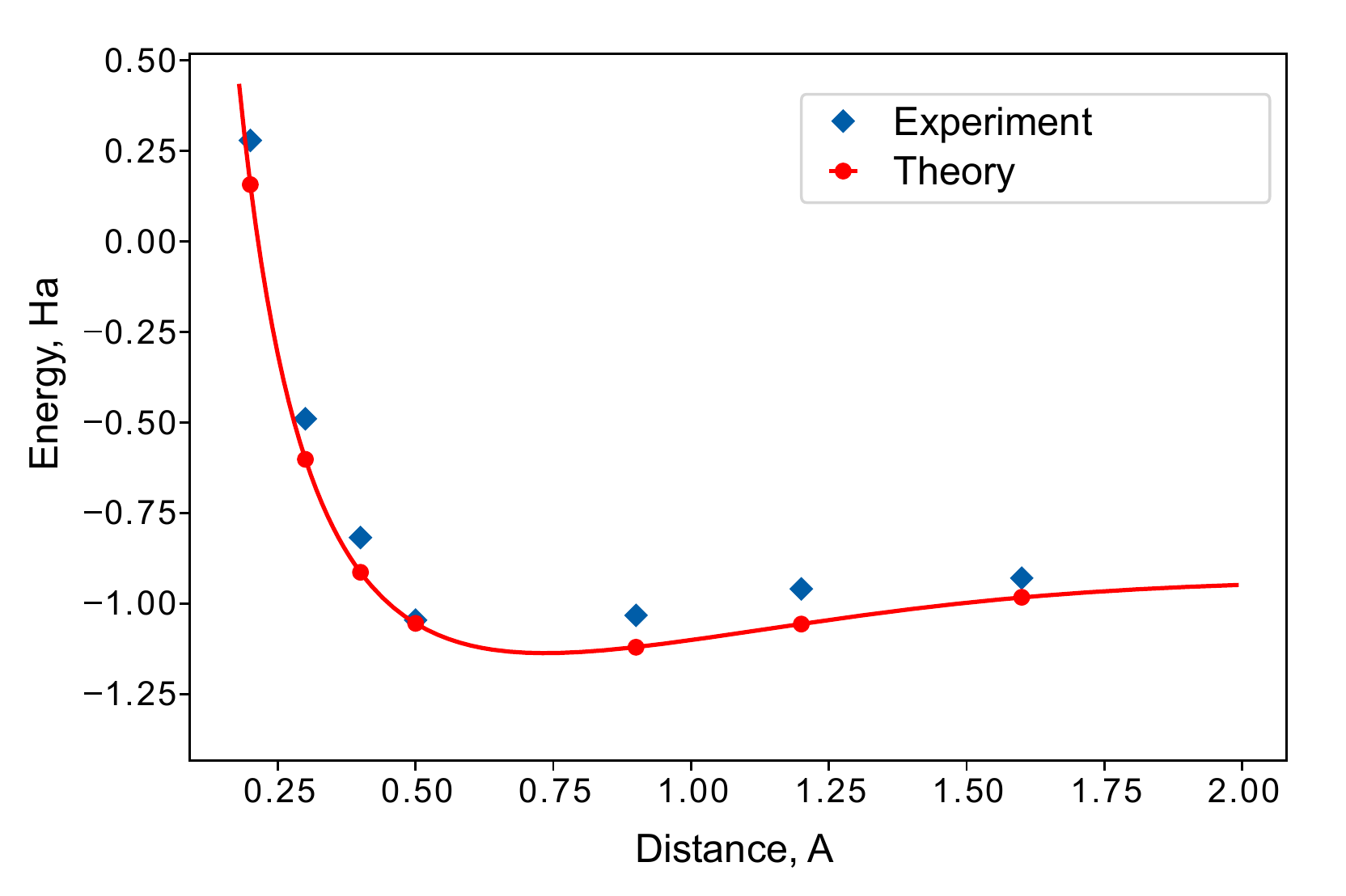}
\caption{\label{fig::H2_vqe} Experimentally measured (blue markers) bound energy of the $H_{2}$ molecule on the two-qubit photonic processor. The red line and the red markers indicate the values of the bound energy computed on a classical processor.}
\end{figure}

The Figure~\ref{fig::H2_vqe}b illustrates the binding energy values corresponding to different internuclei distance calculated using our quantum photonic processor. The red curve and markers show the numerical solution gathered by computing eigenvalues of the Hamiltonian (\ref{eqn:qubit_hamiltonian}).


This simple application illustrates the capability of the processor to deliver performance suitable for small-scale quantum algorithm implementation. The imperfect convergence of the VQE algorithm is mostly due to unstable indistinguishability of the generated photons or not optimal tuning of the optimization routine parameters.

\section{Discussion}

The FLW technology is the unrivaled instrument for rapid prototyping of both passive and reconfigurable integrated photonic circuits for scientific purposes due to low cost and simplicity. We fabricated, characterized and tested the quantum photonic processor using FLW which performs on par with best reconfigurable circuits elaborated in the more sophisticated technological processes based on lithography. We also identify avenues for feasible improvement of the technology.

The loss is the main obstacle in every quantum photonic experiment. The advantage of the FLW developed chips is the naturally low-loss coupling between standard single-mode fibers and the integrated waveguides due to almost identical refractive index contrast of the fiber and the waveguide. Furthermore this feature may be further enhanced by additional tailoring of the refractive index of the waveguide section adjacent to the facet \cite{Heilmann18} and by introducing lensed fiber arrays. The propagation loss can be improved by annealing of the waveguide structure \cite{Bhardwaj04} or by using a multiscan technique \cite{Tan20}. Recent work has demonstrated propagation loss as low as 0.07~dB/cm \cite{Tan22} whereas lithographic technology has reached the loss below 0.001~dB/cm \cite{Bauters11}. 

Another limiting factor specific to FLW technology is low refractive index contrast. Even though this feature enables higher coupling to the fiber without mode converters it significantly reduces the miniaturization potential. The curvature radius can be brought down to 15~mm by using additional suppression walls \cite{Liu2018} or by strong focusing with oil-immersion optics \cite{EATON2011} without drastic increase of propagation loss. however even this radius value will make the circuit with several tens of components a few centimeters long. The long chip inevitably exhibits large total loss and hence the requirements on the propagation loss are more stringent.

The precision of the each individual operation is mainly defined by the quality and reproducibility of the directional couplers. Our technology can be improved significantly by moving the setup to the clean room with temperature stabilised environment. The laser system can also be augmented to deliver more stable pulse width and output power. The optical circuit design is another path towards the robustness of the processor however it may introduce the redundancy in the optical element arrangement \cite{Miller15}. One of the strengths of the FLW fabricated circuits is that they can be postprocessed in order to tailor their properties. For instance additional local exposure helps to tune the coupling ratio of the directional couplers \cite{Will20} which is the unique property of the technology. The FLW fabricated reconfigurable circuits typically feature larger thermal cross-talks. Our results show that a simple algorithm helps to largely exclude the effect and precisely calibrate the phaseshifters. The further improvement includes better thermal isolation \cite{Ceccarelli:19, Ceccarelli2020} and the use of galvanically isolated multichannel current source, which prevents parasitic current flows across the circuit.

The FLW technology may serve as an irreplaceable testbed for prior assessment of novel ideas within a standard optical lab. Even though the FLW-based photonic circuits are subpar to their analogs fabricated in a state-of-the-art nanofabrication facility, the capability to deliver reconfigurable circuits quickly is valuable at the initial stage. The FLW-based photonic platform also allows to add integrated photon sources \cite{Atzeni:18} and detectors \cite{Gerrits2011}.

\section{Conclusion}

Using femtosecond laser writing in a fused silica glass to fabricate an integrated optical chip, we have realized a universal two-qubit quantum photonic processor. To the best of our knowledge, this processor is the most complex and high-quality reconfigurable quantum photonic device for the femtosecond laser written chips up to date. The mean fidelity of the processor's eight single-qubit gates is $F = 99.3\%$ and the fidelity of two-qubit entangling CNOT gate is $F = 94.4\%$. Furthermore, we demonstrated the versatility and potential of the femtosecond laser written chips by estimating the bound energy of the $H_2$ molecule via a VQE experiment based on the fabricated processor.

\begin{acknowledgments}
This work was supported by Rosatom in the framework of the Roadmap for Quantum computing (Contract No. 868-1.3-15/15-2021 dated October 5, 2021 and Contract No.P2154 dated November 24, 2021). S.P.K. is supported by the Ministry of Science and Higher Education of the Russian Federation on the basis of the FSAEIHE SUSU (NRU) (Agreement No. 075-15-2022-1116).
\end{acknowledgments}

\section*{Data Availability Statement}

The data that support the findings of this study are available upon reasonable request.


\nocite{*}
\bibliography{aipsamp}

\appendix

\section{Phaseshifters calibration and cross-talks}\label{app:calibration}

\begin{figure*}[htbp!]
\centering
\includegraphics[width=1.6\columnwidth]{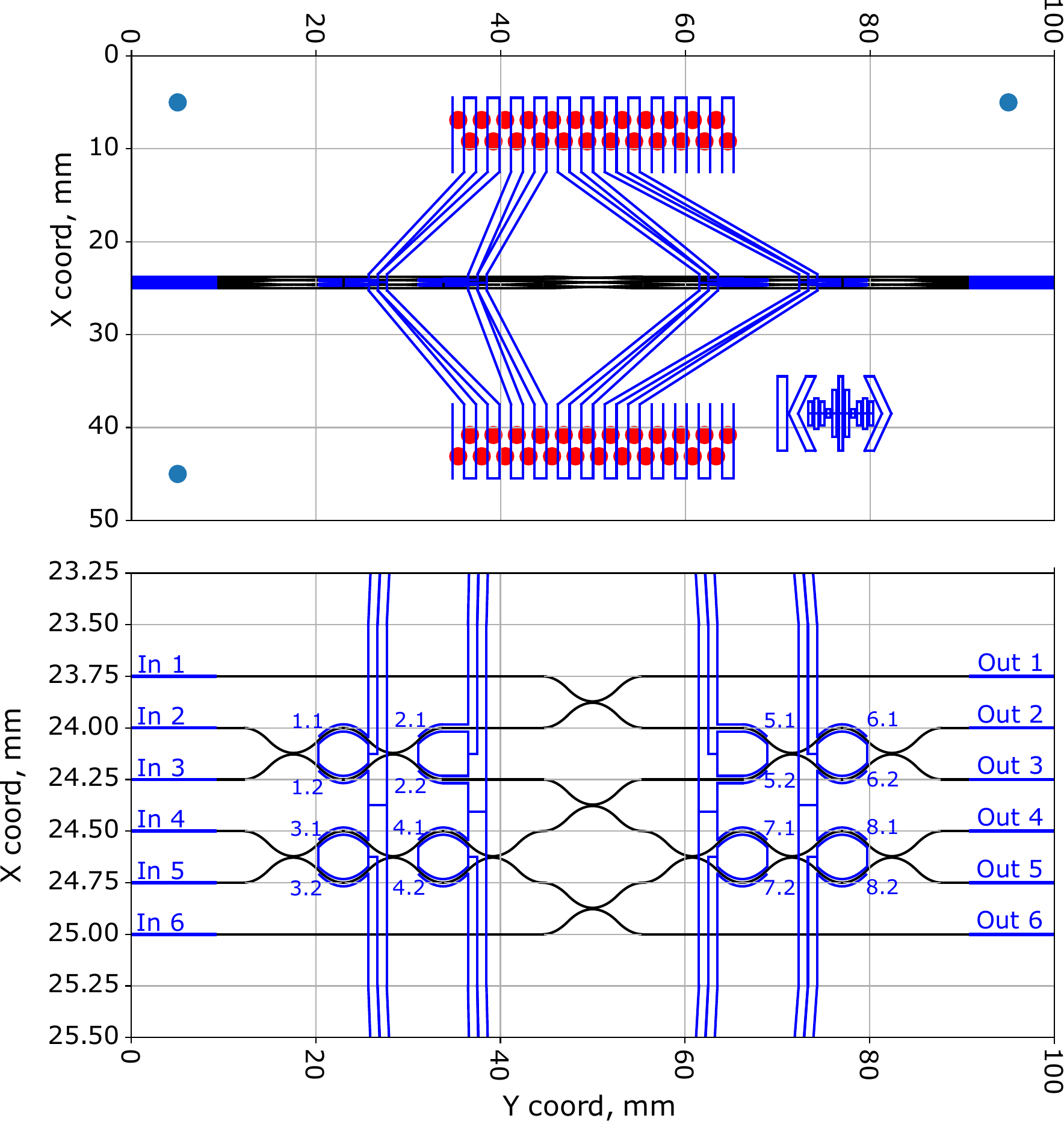}
\caption{\label{fig::chip_structure_act} Actual optical chip structure (view from the top). Up: Real scale. Down: Zoomed. Waveguides are depicted with black lines, engraved electrodes are depicted with blue lines. Red dots illustrate the electrical contacts with PCB. }
\end{figure*}

Schematic structure of the optical chip manufactured by femtosecond laser writing is shown in Fig.~\ref{fig::chip_structure_act}. Each phaseshifter was physically implemented twice by two parallel electrodes to have a backup replacement in case one of them is broken. For example, the logical phaseshifter $\varphi_1$ can be implemented either by electrode $1.1$ or $1.2$ (these two electrodes are duplicates), etc. In the experiment each logical phaseshifter was implemented by a certain single electrode.

In our experiment we observed mutual cross influence of close heaters, which are placed in a vertical column arrangement: the logical heater $1$ (which contains two physical electrodes $1.1$ and $1.2$) affects and is being affected by the logical heater $3$ (which has physical electrodes $3.1$ and $3.2$), the logical heater $2$ affects and is being affected by the logical heater $4$, the logical heater $5$ affects and is being affected by the logical heater $7$, the logical heater $6$ affects and is being affected by the logical heater $8$.
Moreover the heaters cross influence has a sign in the induced phaseshift, which can be determined by a simple rule: the heater $h1$ affects the heater $h2$ with a negative sign if and only if there is a duplicate of the heater $h2$ between the heaters $h1$ and $h2$. This sign rule can be easily understood since duplicated heaters are located in the opposite Mach-Zander interferometer arms. For example, the physical heater $1.2$ affects the physical heater $3.1$ positively and the heater $3.2$ negatively, while both physical heaters $3.1$ and $3.2$ have positive influence on the heater $1.2$. Also no 'horizontal' cross-talk was observed, i.e. the physical heater $1.1$ affects only the heaters $3.1$ and $3.2$ and no others. The same applies to every other physical heater. 

Due to the linear dependence of the phaseshift on squared current, equation (\ref{eqn:phase_crosstalk}) can be written in a matrix form:

\begin{equation}\label{eqn:matrix_equation_crosstalk}
    \Phi = \Phi_0 + A \times I^2,
\end{equation}
where $\Phi = \{\varphi_1, ..., \varphi_n \}$ is a vector of phases to be implemented on chip, $\Phi_0$ is a vector of initial phases which are present on the chip when no current is being applied, $A$ is a cross talk matrix which determines the mutual cross influence of all heaters in the optical chip. The meaning of the cross talk matrix $A = a_{i,j}$ is the influence of $j$-th heater on  $i$-th heater. Thus, diagonal elements of $A$ represent the basic heater action in its own interferometer arm, and non-diagonal elements represent the cross influence of the heaters.
Therefore, to set the desired phases $\Phi = \{\varphi_1, ..., \varphi_n \}$ on the chip one needs to solve the equation (\ref{eqn:matrix_equation_crosstalk}) for the electrical currents which are to be applied. If any of the $I^2_j$ which are the solution of (\ref{eqn:cross-talk_phase_calculation}) appear to be negative, then the corresponding phase $\varphi_j$ in the phase vector $\Phi$ should be increased by $2\pi$. 
The required values of $\Phi_0$ and $A$ are obtained from the calibration procedure which is to be detailed further. 

In the experiment we used the physical heaters $1.1, 2.1, 3.2, 4.2, 5.1, 6.1, 7.2, 8.2$ which correspond to the phases $\varphi_1, \varphi_2, ..., \varphi_8 $ since they are maximally separated along the vertical direction; all the cross talks for these physical heaters are negative.

The heater calibration procedure was performed as follows: classical 808~nm radiation was injected into the particular input mode of the chip, then electrical current was applied to the heater under calibration ranging from $0$ to $20$ mA with the step of $0.15$ mA and the output power was measured in all six output modes of the chip. Finally, the measured data were fitted to obtain the necessary calibration matrix $A$ and the vector of initial phases $\Phi_0$.

\begin{figure}[ht!]
\centering
\includegraphics[width=1.0\columnwidth]{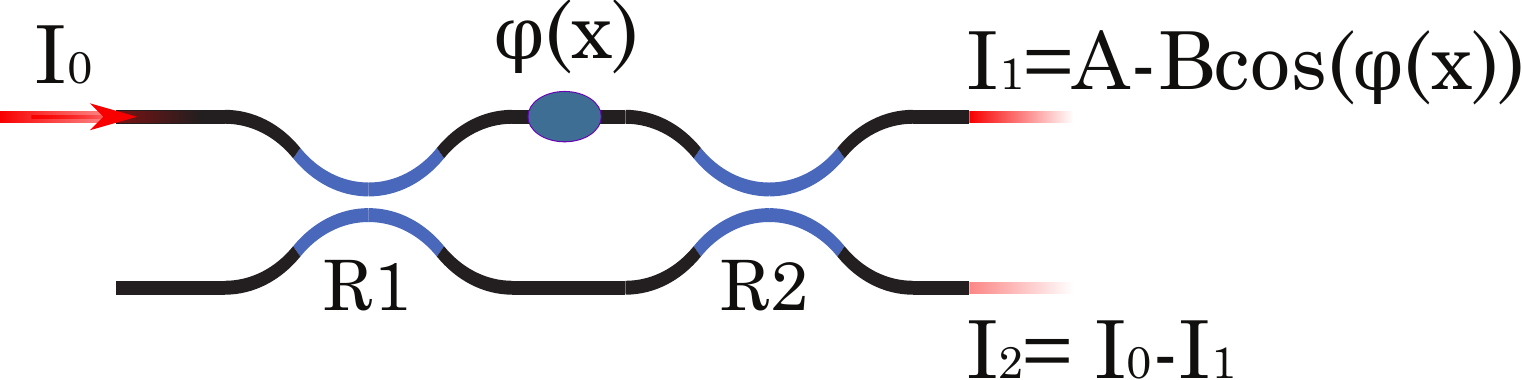}
\caption{\label{fig::mzi_model} Mach-Zander interferometer model used for fitting the calibration curves.}
\end{figure}

We used a Mach-Zander interferometer as a fitting model, since our chip includes four MZIs. The MZI model, shown in Fig.~\ref{fig::mzi_model}, gives the dependence of the output intensity on the applied phase $\varphi = \varphi_0 + \alpha x^2$:

\begin{equation}\label{eqn:fiitng_func_mzi}
    I(x) = B - C \cos(\varphi_0 + \alpha x^2),
\end{equation}
where $I$ is the measured intensity in the selected MZI output mode, $x$ -- the electrical current applied to the heater, $B$, $C$, $\varphi_0$, $\alpha$ are the fitting parameters. Expression (\ref{eqn:fiitng_func_mzi}) is valid for a Mach-Zander interferometer consisting of two integrated directional couplers with power reflectivities $R_1$ and $R_2$, as depicted in Fig [\ref{fig::mzi_model}].  
In addition, the $B$ and $C$ coefficients may be expressed via the power reflectivities $R_1$ and $R_2$:
\begin{gather*}\label{eqn:fiitng_func_B_C}
   B = 1 - (R_1 + R_2) + R_1 R_2, \\
   C = 2 \sqrt{ R_1 R_2 (1 - R_1) (1 - R_2) },
\end{gather*}
and implicit information about $R_1$ and $R_2$ can be obtained from the $B$ and $C$ coefficients.

Calibration data were gathered as follows: 

\begin{enumerate}
  \item Laser radiation was injected in the first input mode (see Fig.~\ref{fig::chip_structure_act}) and current up to $20$~mA was applied to the heater $6$ ($6.1$ or $6.2$ depending on what electrode was chosen) and to the heater $8$ ($8.1$ or $8.2$) to measure its cross-talk influence on heater $6$. Two power dependencies corresponding to the MZI model in this measurement are modes $2$ and $3$ (see Fig.~\ref{fig::chip_structure_act}).
  
  \item Then coherent laser radiation was launched in the second input mode and current was applied to the heater $1$ as well as heater $3$ to measure its cross-talk influence on heater $1$. Power dependencies used for fitting were the output mode $1$ and sum of the output modes $4$ and $5$.
  
  \item After that $\pi/2$ phases were set at heaters $1$ and $6$, currents were applied to heaters $2$ and $5$, and their cross-talk neighbours --- heaters $4$ and $7$. Power dependencies used for fitting were the output modes $2$ and $3$.
  
  \item Next, radiation was injected in the $4$-th input mode and currents were applied to heaters $3$, $4$ and to their crosstalk neighbours heaters $1$ and $2$ respectively. Output modes used for fitting were the sum of output modes $2$ and $3$ and output mode $6$. Here we should note that $\pi/2$ phase should be set with the heater $3$ while calibrating the heater $4$ and vice versa to achieve the best visibility of the output interference fringes.
  
  \item Finally, input radiation was launched in the $6$-th input mode and currents were applied to the heaters $7$, $8$ and to their cross-talk neighbours -- heaters $5$ and $6$ respectively. Power dependencies used for fitting were the output modes $4$ and $5$. Again, we should note that $\pi/2$ phase should be set with heater $7$ while calibrating heater $8$ and vice versa to achieve the best visibility of the output interference fringes.
  
\end{enumerate}

\begin{figure*}[htbp!]
\centering
\includegraphics[width=2\columnwidth]{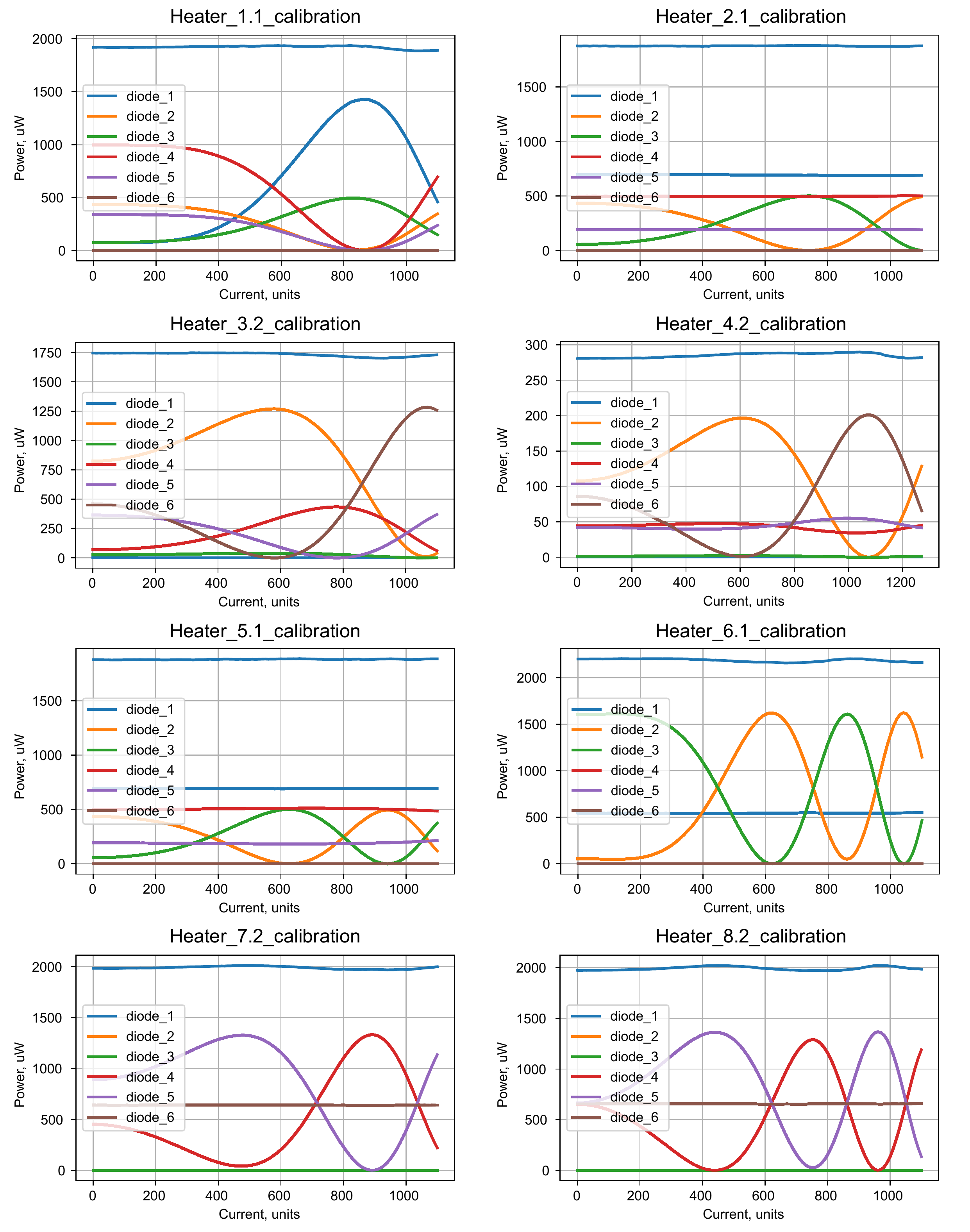}
\caption{\label{fig::calibration_data_fig} Heaters calibration data obtained from the measurements performed with classical coherent input light at 808~nm wavelength. Current is in relative units, 1000 of which corresponds to 15 mA and 0 units means zero current. The upmost horizontal blue line in each plot demonstrates the sum over all output powers.}
\end{figure*}

After all necessary calibration data were measured, the data were fitted according to the [\ref{eqn:fiitng_func_mzi}] and necessary calibration entities $A$ and $\Phi_0$ were revealed. The values of $A$ and $\Phi_0$ elements are listed in Table \ref{tab::crosstalk_matrix}.

\begin{table}[ht!]
    \centering
    \begin{tabular}{ |c|c|c|c|c|c|c|c|c| } 
     \hline
     htr No & 1.1 & 2.1 & 3.2 & 4.2 & 5.1 & 6.1 & 7.2 & 8.2 \\ \hline
     1.1 & 4.37 & 0    & -0.71 & 0    & 0    & 0    & 0    & 0    \\ \hline
     2.1 & 0    & 4.49 & 0    & -0.78 & 0    & 0    & 0    & 0    \\ \hline
     3.2 & -0.73 & 0    & 4.61 & 0    & 0    & 0    & 0    & 0    \\ \hline
     4.2 & 0    & -0.70 & 0    & 4.43 & 0    & 0    & 0    & 0    \\ \hline
     
     5.1 & 0    & 0    & 0    & 0    & 4.64 & 0     & -0.82 & 0    \\ \hline
     6.1 & 0    & 0    & 0    & 0    & 0    & 4.90  &  0    & -0.85 \\ \hline
     7.2 & 0    & 0    & 0    & 0    & -0.66 & 0    & 4.61 & 0    \\ \hline
     8.2 & 0    & 0    & 0    & 0    & 0    & -0.83 & 0    & 5.21 \\
     \hline
     \hline
     $\Phi_0$ & -0.20 & -0.01 & -0.05 & 0.15 & -0.01 & -0.21 & 0.250 & 0.285 \\ \hline
    \end{tabular}
    \caption{\label{tab::crosstalk_matrix} The cross-talk matrix $A$ for the heaters and vector $\Phi_0$ of initial phase shifts in optical chip obtained from calibration procedure. All elements of $A$ are expressed in $10^{-2}$ rad/mA$^2$, and elements of $\Phi_0$ are expressed in radians. }
\end{table}

\section{ Photons indistinguishability tuning }\label{app:photons_indist_tune}

Before implementing an applied quantum experiment with single photons on chip their mutual indistinguishability should be adjusted. 

In our setup pairs of single photons are generated from an SPDC source (see Fig.~\ref{fig::ExpSetup}) and controllable experimental parameters affecting the photons mutual indistinguishablility are: PPKTP crystal temperature affecting the wavelength of generated photon pairs, HWP and QWP plates angles in the SPDC source controlling the polarization of the photons, and the position of the linear translation system (LTS) which is used to compensate the optical path difference for two photons.

\begin{figure*}[ht!]
\centering
\includegraphics[width=1.2\columnwidth]{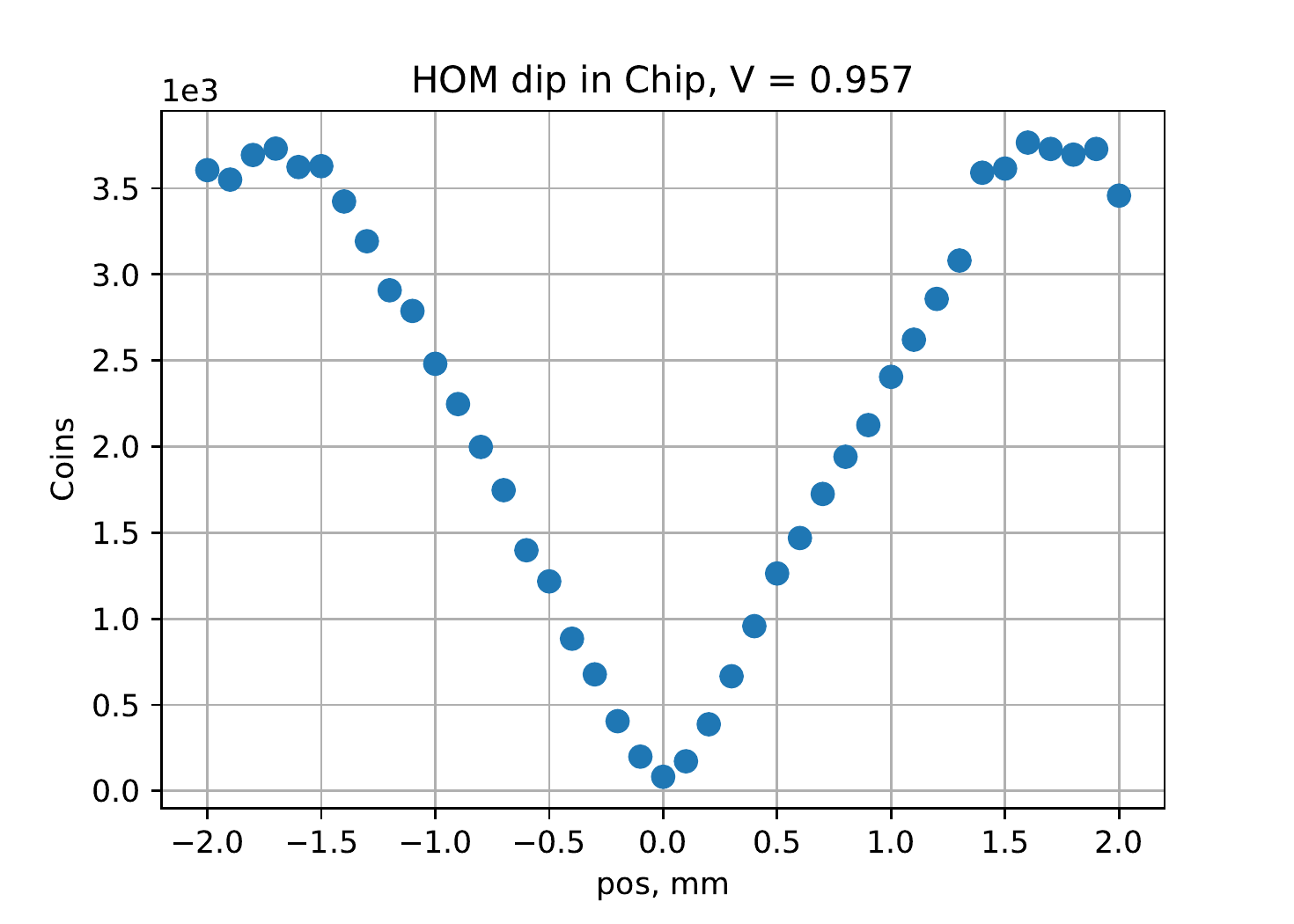}
\caption{\label{fig::hom_dips} Experimentally measured HOM interference inside the chip.}
\end{figure*}

Before injecting the SPDC photon pair into the chip we launched them into the $50:50$ fiber beam splitter and adjusted the necessary experimental parameters to maximize their mutual indistinguishability which was estimated by measuring the Hong-Ou-Mandel (HOM) dip in coincident photon detection counts from both output channels of the fiber beam splitter. Temperature of the SPDC crystal corresponding to the degenerate regime of photon pairs generation was found to be around $37^\circ C$ and could fluctuate with the temperature in the laboratory room. Then, one of the output fiber coupler positioned on the linear translation system (LTS) was moved and coincident detection counts between the corresponding channels were measured. Finally we ran a similar test using the chip in a specific configuration as a balanced beamsplitter and got the HOM interference fringe illustrated in Fig.~\ref{fig::hom_dips}. HOM-dip in chip was obtained with special currents set on phase shifters such that the whole chip transformation was equal to the CNOT gate (all single qubit gates were equal to identity transformation); photons were injected into third and fourth input modes of the chip and dip in coincidences was observed between third and fifth output modes of the chip.

\section{Quantum process tomography of two qubit gate}\label{app:qpt_appdx}

As it is known a quantum process $\mathcal{E}$ links an input state $\rho$ with an output state $\rho_{out} = \mathcal{E} (\rho)$, where $\rho$ is a state density matrix. Such quantum process can be described by a completely positive map:

\begin{equation}\label{eqn:Q_Process}
    \mathcal{E} (\rho) = \sum_{i} A_i \rho A_{i}^{\dagger},
\end{equation}

where elements $A_i$ satisfy  $\sum_{i} A_{i}^{\dagger} A_i \leq I $ which in turn leads to $Tr \left[ \mathcal{E} (\rho) \right] \leq 1 $ condition. 

If ${E_i}$ is an orthogonal basis in a Hilbert space in which $\rho$ is defined, then elements $A_i$ may be decomposed as $ A_i = \sum_{m} a_{i m} E_m$ which finally yields the quantum process $\mathcal{E}$ representation via $\chi$-matrix:

\begin{equation}\label{eqn:Q_Process_and_chi}
    \mathcal{E} (\rho) = \sum_{m, n} \chi_{mn} E_{m} \rho E_{n}^{\dagger},
\end{equation}
with $\chi_{m n} = \sum_{i} a_{i m} a_{i n}^{\dagger} $.
The main aim of the quantum process tomography is to find $\chi$-matrix, which is a positive superoperator and completely characterizes $\mathcal {E}$  with respect to the $E_{i}$ basis.
We chose Pauli matrices as an orthogonal basis for Hilbert space.

To experimentally obtain the $\chi$-matrix we followed the standard quantum process tomography routine \cite{chuang1997prescription}, which suggests to send into the $N$-qubit gate under characterization a set of $4^N$ linearly independent known states $\rho_{in}$ and for each input state conduct a set of $4^N$ projective measurements in linearly independent bases (quantum state tomography), resulting in making $4^N \times 4^N $ measurements. 

After obtaining experimental data the maximum likelihood procedure was used to seek for an optimal process $\chi$-matrix which fits the measurement results the best. 

During the optimization the cost function was:
\begin{equation}\label{eqn:qpt_cost_function}
    F(t) = \sum \left[ P_{theory}(t) - P_{experiment}  \right]^2,
\end{equation}
where $t$ are the optimization parameters, $ P_{theory} (t) $ - predicted value from model with parameter set $t$, $ P_{experiment} $ - value measured in experiment. Physical meaning of the value $P$ is a probability of measuring output state $\rho_{out}$ from a gate described by a set of parameters $t$ in a basis state $\tau$:
\begin{equation}\label{eqn:qpt_prob_model}
    P(t) = Tr[ \tau \times \rho_{out}(t) ] = Tr[ \tau \times  \mathcal{E} (\rho_{in}) ],
\end{equation}
and for $\rho_{out}$:
\begin{equation}\label{eqn:qpt_paramterization_chi}
    \rho_{out} = \sum_{m, n} \chi_{mn}(t) E_{m} \rho_{in} E_{n}^{\dagger},
\end{equation}
where $\chi_{m,n}(t)$-process matrix to be determined parameterized with $4^N \times 4^N $ real parameters $t$. Here input states $\rho_{in}$, projective states $\tau$ and two qubit Pauli matrices $A_i$ are assumed to be known and therefore value of $ P_{theory} (t)$ can be calculated and compared with corresponding experimentally measured value $ P_{experiment} $.

In our case for two qubit CNOT gate $N=2$ and therefore $\chi$-matrix has dimensions $16\times 16$ and requires 256 projective measurements. 


The $\chi$-matrix hypothesis was parameterized with 256 real numbers $\Vec{t}$ according to \cite{altepeter2005photonic}:
\begin{equation}\label{eqn:chi_parametrization}
\chi(\Vec{t}) =  \frac{g(\Vec{t})g(\Vec{t})^{\dagger}}{Tr[ g(\Vec{t})g(\Vec{t})^{\dagger} ]},
\end{equation}
where $g(\Vec{t})$ can be an arbitrary complex matrix parametrization, for example, the triangular one \cite{altepeter2005photonic}. The optimization was carried out using the SciPy package.

In experiment two single indistinguishable photons from SPDC source were injected into the second and fourth input modes of the chip and four two-fold coincidence counts  were detected: $C_1$ - coincidence detection between output modes №2 and №4, $C_2$ - between output modes №2 and №5, $C_3$ - between output modes №3 and №4 and $C_4$ - between output modes №3 and №5, which corresponds to detection of two output qubits in the logical states $\ket{00}$, $\ket{01}$, $\ket{10}$ and $\ket{11}$.
Values of $ P_{experiment}(\rho_{out}, \tau) $ were determined as:
\begin{equation}\label{eqn:qpt_prob_measure}
P_{experiment}(\rho_{out}, \tau) = \frac{C_{\tau}}{ \sum_{i} C_i },
\end{equation}
where $C_{\tau}$ is a number of counts corresponding to projective measurement of $\tau$ state. Additionally, directly measured counts $C_{1}$ - $C_{4}$ were then multiplied by four real numbers $e_1 - e_4$ respectively, which has the meaning of correction for different detection efficiencies. The procedure of obtaining four coincidence detection efficiencies $e_1, e_{2}, e_{3}, e_4$ was the following:

\begin{enumerate}
  \item Two qubit state $\ket{10}$ was injected into the CNOT gate. The input single qubit gates $R_{x1}, R_{z1}, R_{x2}, R_{z2}$ prepared the required  $\ket{10}$ state. The CNOT gate operation results in the output two qubit state $\ket{10}$. In experiment only $C_{3}$ coincidence configuration must be registered.
  
  \item Then using the output single qubit gates $R_{x3}, R_{z3}, R_{x4}, R_{z4}$ output two qubit state $\ket{10}$ was transformed into $\ket{00}$, $\ket{01}$ and $\ket{11}$, which correspond to detecting coincidence counts only in $C_{1}$, $C_{2}$ and $C_{4}$ configurations respectively.
  
  \item Assuming all the measured counts $C_1, C_2, C_3, C_4$ must be equal as they share the same input two photon source the values of detection efficiencies $e_1, e_2, e_3, e_4$ can be evaluated:
\end{enumerate}
\begin{equation}\label{eqn:qpt_efficiences}
    C_i e_i = const, \ \  i = 1,2,3,4.
\end{equation}

Initially input two qubit state injected into the chip was $\ket{00}$ and it is manipulated with the help of the input single qubit gates $R_{x1}, R_{z1}, R_{x2}, R_{z2}$. The projective measurements of two output qubits were performed with the help of output single qubit gates $R_{x3}, R_{z3}, R_{x4}, R_{z4}$. The following set of states 
was used both as the input to the CNOT gate and as the set of states for projective measurements:
\begin{gather*}\label{eqn:qpt_qubit_states}
  \ket{H} = \ket{0}, \ \ket{V} =  \ket{1}, \\
  \ket{D} = \frac{\ket{0} + \ket{1}}{\sqrt{2}}, \  
  \ket{A} = \frac{\ket{0} - \ket{1}}{\sqrt{2}}, \\
  \ket{R} = \frac{\ket{0} + i \ket{1}}{\sqrt{2}}, \ 
  \ket{L} = \frac{\ket{0} - i \ket{1}}{\sqrt{2}}
\end{gather*}
Here labels H, V, A, D, R and L are used only as an analogy to polarizational states for convenience of writing. Actual qubit states in the experiment were in dual-rail encoding.

To prepare each input single qubit state from the above set phases listed in Table \ref{tab::qpt_input_state_phases} were set on chip. 
These phases should be set on the corresponding input phaseshifters $\varphi_1, \varphi_2$ and $\varphi_3, \varphi_4$ for the first and second qubit respectively (see Fig. \ref{fig::chip_structure_act}). For example, if one needs to prepare $\ket{DL}$ state, which means first qubit is in $\ket{D}$ state and second qubit is in $\ket{L}$ state, following phases on chip should be set: $\varphi_1 = \pi/2, \varphi_2  =  \pi/2$ and $\varphi_3 = \pi/2, \varphi_4 = 0$.
\begin{table}[h]
    \centering
    \begin{tabular}{ |c|c|c| } 
     \hline
     State & phase 1 & phase 2 \\ \hline
     $\ket{H}$ & $\pi$ & $\pi$  \\ \hline
     $\ket{V}$ & 0 & 0  \\ \hline
     $\ket{D}$ & $\pi/2$ & $\pi/2$  \\ \hline
     $\ket{A}$ & $\pi/2$ & $3\pi/2$  \\ \hline
     $\ket{R}$ & $\pi/2$ & $\pi$  \\ \hline
     $\ket{L}$ & $\pi/2$ & 0  \\ \hline
    \end{tabular}
    \caption{\label{tab::qpt_input_state_phases} Phases needed to be set on chip to prepare particular input single qubit state. }
\end{table}


The phase settings listed below were used for projective measurements in particular basis:

\begin{table}[h]
    \centering
    \begin{tabular}{ |c|c|c| } 
     \hline
     Basis & phase 1 & phase 2 \\ \hline
     HV & $\pi$ & $\pi$  \\ \hline
     DA & $\pi/2$ & $\pi/2$  \\ \hline
     RL & 0 & $\pi/2$  \\ \hline
    \end{tabular}
    \caption{\label{tab::qpt_output_state_phases} Phases needed to be set on chip for projective measurements in particular basis. }
\end{table}

These phases should be set on the corresponding output phase shifters $\varphi_5, \varphi_6$ and $\varphi_7, \varphi_8$ for the first and second qubit respectively (see Fig. \ref{fig::chip_structure_act}). For example, if one needs to measure first qubit in RL basis and second qubit in HV basis, following phases on chip should be set: $\varphi_5 = 0, \varphi_6  =  \pi/2$ and $\varphi_7 = \pi, \varphi_8 = \pi$.

Experimentally obtained data counts are presented in Table \ref{tab::qpt_data_coins}. Each row in Table \ref{tab::qpt_data_coins} was a single experimental run of a type: "input state preparation to CNOT and projective measurement of output state from CNOT". First column contains information encoded in four letters about the particular experimental run: first two capital letters encode input two qubit state which was sent to a quantum process and the last two letters encode the bases of the projective measurement. Next four columns hold measured two-fold coincidence counts ($C1$ - $C4$), and the last column includes the sum of all counts of a particular experimental run $\sum C_j$. For example, configuration "VDrd" means that input two qubit state in CNOT was $\ket{V} \times \ket{D}$, which is first qubit was in $\ket{V}$ state and second qubit was in $\ket{D}$ state, and projective measurements were performed in $ r=\ket{R}, \ket{L}$ basis for the first qubit and in $ d=\ket{D}, \ket{A}$ basis for the second qubit; thus, listed two-fold coincidence counts in this row correspond to measured projections of output state $\rho_{out}$ of the CNOT onto following states $\tau$: $C_{1}$ - $\tau = \ket{R} \times \ket{D}$, $C_{2}$ - $\tau = \ket{R} \times \ket{A}$, $C_{3}$ - $\tau = \ket{L} \times \ket{D}$, $C_{4}$ - $\tau = \ket{L} \times \ket{A}$. Analogously, configuration "LAhr" stands for $\ket{L} \times \ket{A}$ input two qubit state in CNOT, and measured counts correspond to  projections onto states $\tau$: $C_{1}$ - $\tau = \ket{H} \times \ket{R}$, $C_{2}$ - $\tau = \ket{H} \times \ket{L}$, $C_{3}$ - $\tau = \ket{V} \times \ket{R}$, $C_{4}$ - $\tau = \ket{V} \times \ket{L}$.

As noted earlier, to fully characterize the process $\chi-$matrix of the CNOT gate, at least 256 independent projection measurements must be carried out. One experimental run (one row in Table \ref{tab::qpt_data_coins}) "preparation of the input state in the CNOT gate and projection measurement of the output state from the CNOT gate" simultaneously produces four projection measurements: $C_{1}$ to $C_{4}$. Therefore, for quantum process tomography of the CNOT gate, it is necessary to conduct at least 64 experimental measurement runs with 64 independent configurations.

After carrying out the necessary measurements, we ran the optimization of the cost function (\ref{eqn:qpt_cost_function}) over the $\chi$-matrix parameters. The result is the $\chi$-matrix of the process of the CNOT gate presented in the main text in Fig. \ref{fig::gates_fidelity}.

\setlength\tabcolsep{13pt}
\begin{table*}[htbp!]
    \centering
    \begin{tabular}{ |c|c|c|c|c|c| } 
        \hline
        Config & C1 & C2 & C3 & C4 & Sum \\ \hline
        HHhh  & 13    & 2258  & 1     & 0     & 2270 \\ \hline
        HVhh  & 1914  & 39    & 16    & 4     & 1973 \\ \hline
        VHhh  & 0     & 15    & 2095  & 248   & 2358 \\ \hline
        VVhh  & 30    & 1     & 337   & 2357  & 2725 \\ \hline
        HRhr  & 101   & 1898  & 4     & 11    & 2014 \\ \hline
        HLhr  & 1823  & 52    & 12    & 0     & 1887 \\ \hline
        VRhr  & 4     & 22    & 1965  & 366   & 2358 \\ \hline
        VLhr  & 26    & 7     & 326   & 2185  & 2543 \\ \hline
        HDhd  & 1762  & 7     & 27    & 0     & 1793 \\ \hline
        HAhd  & 4     & 2063  & 0     & 7     & 2074 \\ \hline
        VDhd  & 13    & 0     & 2467  & 29    & 2509 \\ \hline
        VAhd  & 0     & 21    & 119   & 1850  & 1985 \\ \hline
        DDdd  & 2025  & 7     & 176   & 15    & 2223 \\ \hline
        DAdd  & 0     & 17    & 35    & 1945  & 1995 \\ \hline
        ADdd  & 174   & 10    & 1858  & 3     & 2044 \\ \hline
        AAdd  & 50    & 1973  & 3     & 15    & 2041 \\ \hline
        RDrd  & 2106  & 0     & 182   & 14    & 2303 \\ \hline
        RArd  & 7     & 15    & 35    & 1876  & 1933 \\ \hline
        LDrd  & 199   & 7     & 1915  & 6     & 2127 \\ \hline
        LArd  & 50    & 1867  & 1     & 36    & 1954 \\ \hline
        HHhd  & 973   & 951   & 0     & 0     & 1916 \\ \hline
        HVhd  & 1119  & 812   & 0     & 0     & 1931 \\ \hline
        VHhd  & 9     & 10    & 1196  & 981   & 2195 \\ \hline
        VVhd  & 7     & 14    & 1753  & 566   & 2339 \\ \hline
        HHhr  & 1370  & 523   & 3     & 0     & 1891 \\ \hline
        HVhr  & 693   & 1263  & 16    & 3     & 1975 \\ \hline
        VHhr  & 10    & 6     & 663   & 1635  & 2313 \\ \hline
        VVhr  & 18    & 17    & 1370  & 964   & 2369 \\ \hline
        HDhh  & 718   & 1053  & 14    & 9     & 1794 \\ \hline
        HAhh  & 815   & 1039  & 3     & 3     & 1860 \\ \hline
        VDhh  & 4     & 12    & 1377  & 1226  & 2620 \\ \hline
        VAhh  & 11    & 18    & 817   & 846   & 1692 \\ \hline
        HDhr  & 943   & 883   & 0     & 11    & 1832 \\ \hline
        HAhr  & 869   & 1060  & 0     & 4     & 1933 \\ \hline
        VDhr  & 20    & 6     & 1279  & 1240  & 2545 \\ \hline
        VAhr  & 4     & 14    & 801   & 1027  & 1846 \\ \hline
        HRhh  & 613   & 1127  & 3     & 5     & 1748 \\ \hline
        HLhh  & 826   & 834   & 0     & 7     & 1662 \\ \hline
        VRhh  & 6     & 12    & 1163  & 877   & 2058 \\ \hline
        VLhh  & 24    & 6     & 920   & 1079  & 2029 \\ \hline
        HRhd  & 796   & 801   & 5     & 8     & 1611 \\ \hline
        HLhd  & 804   & 732   & 7     & 10    & 1553 \\ \hline
        VRhd  & 7     & 12    & 1163  & 793   & 1976 \\ \hline
        VLhd  & 7     & 11    & 1237  & 669   & 1924 \\ \hline
        HDdd  & 734   & 1     & 692   & 5     & 1433 \\ \hline
        HAdd  & 3     & 875   & 6     & 754   & 1638 \\ \hline
        VDdd  & 1148  & 11    & 1147  & 11    & 2316 \\ \hline
        VAdd  & 30    & 703   & 76    & 901   & 1710 \\ \hline
        HDrd  & 822   & 3     & 692   & 5     & 1522 \\ \hline
        HArd  & 4     & 923   & 4     & 810   & 1742 \\ \hline
        VDrd  & 1088  & 8     & 1379  & 16    & 2492 \\ \hline
        VArd  & 37    & 989   & 25    & 777   & 1829 \\ \hline
        HHdh  & 11    & 999   & 11    & 763   & 1784 \\ \hline
        HVdh  & 895   & 7     & 661   & 19    & 1581 \\ \hline
        VHdh  & 814   & 102   & 949   & 101   & 1966 \\ \hline
        VVdh  & 158   & 1005  & 131   & 1024  & 2317 \\ \hline
        HHrh  & 21    & 908   & 15    & 681   & 1625 \\ \hline
        HVrh  & 818   & 19    & 674   & 13    & 1525 \\ \hline
        VHrh  & 853   & 103   & 940   & 131   & 2028 \\ \hline
        VVrh  & 162   & 886   & 211   & 1135  & 2394 \\ \hline
        RRrd  & 920   & 7     & 92    & 855   & 1874 \\ \hline
        RLrd  & 1045  & 12    & 99    & 733   & 1890 \\ \hline
        LRrd  & 146   & 906   & 973   & 19    & 2044 \\ \hline
        LLrd  & 124   & 857   & 1025  & 6     & 2012 \\ \hline
    \end{tabular}
    \caption{\label{tab::qpt_data_coins} Measured data for quantum process tomography of two qubit CNOT gate. }
\end{table*}

\section{Chip operation simulation}\label{app:chip_simul}
We developed a numerical engine to simulate the operation of the investigated two-qubit processor. We used the simulation results to analyze experimentally extracted $\chi$-matrix and infer the knowledge about intrinsic defects of the processor.

\begin{figure*}[ht!]
\centering
\includegraphics[width=1.5\columnwidth]{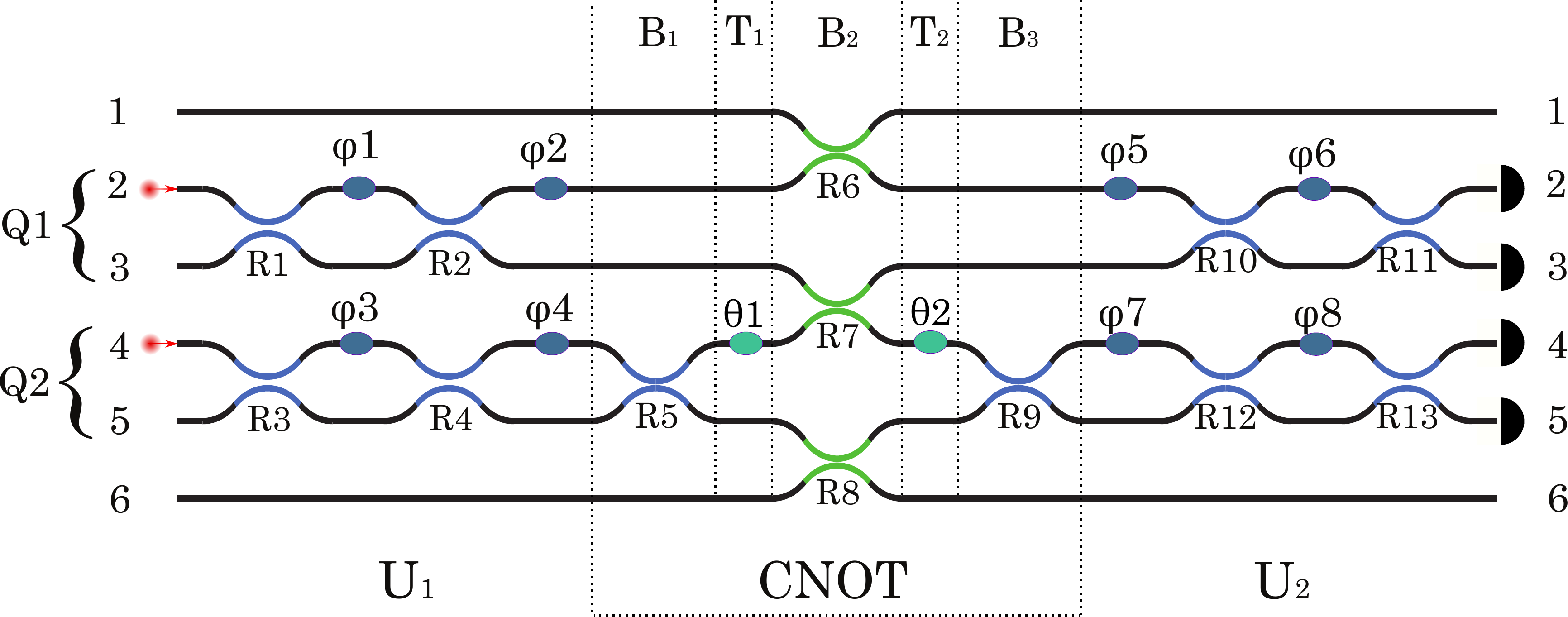}
\caption{\label{fig::simScheme} Schematic image of the two qubit chip waveguide structure used for simulation.}
\end{figure*}

Unitary transformation corresponding to the chip depicted in figure \ref{fig::simScheme} can be defined as:
\begin{equation}
U = U_2 \times CNOT \times U_1,
\end{equation}
where
\begin{equation*}
U_{1,2} = 
\begin{pmatrix}
1 & 
\begin{matrix}
0 & 0
\end{matrix}
& \begin{matrix}
0 & 0
\end{matrix} & 0 \\

\begin{matrix}
0 \\ 0
\end{matrix} 
& MZI_1 & 
\begin{matrix}
0 & 0 \\ 0 & 0
\end{matrix}  & 
\begin{matrix}
0 \\ 0
\end{matrix} \\

\begin{matrix}
0 \\ 0
\end{matrix}  & 
\begin{matrix}
0 & 0 \\ 0 & 0
\end{matrix}  & 
MZI_2 & \begin{matrix}
0 \\ 0
\end{matrix}  \\

0 & \begin{matrix}
0 & 0
\end{matrix} & 
\begin{matrix}
0 & 0
\end{matrix} &
1 
\end{pmatrix},
\end{equation*}

\begin{equation}
MZI_{1,2} = DC(R_2) \times P(\phi) \times DC(R_1),
\end{equation}

\begin{equation}
DC(R) = 
\begin{pmatrix}
\sqrt{R} & i \sqrt{1 - R} \\
i \sqrt{1 - R} & \sqrt{R}
\end{pmatrix},
\end{equation}

\begin{equation}
P(\phi) = 
\begin{pmatrix}
\exp{i \phi} & 0 \\
0 & 1
\end{pmatrix},
\end{equation}

\begin{equation}
CNOT = B_3 T_2 B_2 T_1 B_1,
\end{equation}
where

\begin{equation*}
B_{1,3} = 
\begin{pmatrix}
1 & 0 & 0 &
\begin{matrix}
0 & 0
\end{matrix} & 0 \\

0 & 1 & 0 &
\begin{matrix}
0 & 0
\end{matrix} & 0 \\

0 & 0 & 1 &
\begin{matrix}
0 & 0
\end{matrix} & 0 \\

\begin{matrix}
0 \\ 0
\end{matrix} & 
\begin{matrix}
0 \\ 0
\end{matrix} &
\begin{matrix}
0 \\ 0
\end{matrix} &
DC(R) &
\begin{matrix}
0 \\ 0
\end{matrix} \\

0 & 0 & 0 &
\begin{matrix}
0 & 0
\end{matrix} & 1 \\
\end{pmatrix},
\end{equation*}

\begin{equation*}
T_{1,2} = 
\begin{pmatrix}
1 & 0 & 0 &
\begin{matrix}
0 & 0
\end{matrix} & 0 \\

0 & 1 & 0 &
\begin{matrix}
0 & 0
\end{matrix} & 0 \\

0 & 0 & 1 &
\begin{matrix}
0 & 0
\end{matrix} & 0 \\

\begin{matrix}
0 \\ 0
\end{matrix} & 
\begin{matrix}
0 \\ 0
\end{matrix} &
\begin{matrix}
0 \\ 0
\end{matrix} &
P(\theta) &
\begin{matrix}
0 \\ 0
\end{matrix} \\

0 & 0 & 0 &
\begin{matrix}
0 & 0
\end{matrix} & 1 \\
\end{pmatrix},
\end{equation*}

\begin{equation}
B_2 = 
    \begin{pmatrix}
    DC(R_1) &
    \begin{matrix}
        0 & 0 \\
        0 & 0
    \end{matrix} &
    \begin{matrix}
        0 & 0 \\
        0 & 0
    \end{matrix} \\
    
    \begin{matrix}
        0 & 0 \\
        0 & 0
    \end{matrix} & DC(R_2) &
    \begin{matrix}
        0 & 0 \\
        0 & 0
    \end{matrix} \\ 
    \begin{matrix}
        0 & 0 \\
        0 & 0
    \end{matrix} &
    \begin{matrix}
        0 & 0 \\
        0 & 0
    \end{matrix} & DC(R_3)
    \end{pmatrix}.
\end{equation}

Therefore, the unitary matrix describing the chip operation has 23 real parameters: 13 power splitting ratios $R_1 \ldots R_{13}$, eight tunable phase shifts $\varphi_1 \ldots \varphi_8$ and 2 static phase shifts $\theta_1, \theta_2$. In case of an ideal chip $R_{6,7,8} = 0.33$, all other $R_{j\neq 6,7,8} = 0.5$ and $\theta_{1,2} = 0$. 

The photon Fock state at the chip input was $\ket{0, 1, 0, 1, 0, 0}$ and two-photon coincidences registered in the experiment correspond to the following Fock states: $C_1$ - $\ket{0, 1, 0, 1, 0, 0}$, $C_2$ - $\ket{0, 1, 0, 0, 1, 0}$, $C_3$ - $\ket{0, 0, 1, 1, 0, 0}$, $C_4$ - $\ket{0, 0, 1, 0, 1, 0}$.

In case of $n$ indistinguishable photons the probability of measuring the output Fock state $\ket{\mathrm{Out}} = \ket{j_1, j_2, \dots, j_m}$ for the input Fock state $\ket{\mathrm{In}} = \ket{i_1, i_2, \dots, i_m}$ propagating through the optical chip with $m$ modes described by the unitary matrix $U$ can be calculated as \cite{Tillmann2013}:

\begin{equation}
\label{prob_indist}
    P_j = \frac{|\mathrm{Perm}(U_{in,out})|^2}{i_1!i_2! \dots i_m!j_1!j_2! \dots j_m! },
\end{equation}
where $\mathrm{Perm}(U_{in,out})$ is the permanent of the matrix $U_{in,out}$, which can be constructed using $U$, $\ket{In}$, and $\ket{Out}$.

For two partly indistinguishable photons with the state vectors $\ket{\psi_{1,2}}$ the formula for the probability of measurement outcomes becomes \cite{tichy2015sampling}:

\begin{equation}
\label{prob_disting}
    P_j = \sum_{\sigma \in \Sigma} \left( \prod_{j=1}^2 S_{\sigma_j j}  \right) \mathrm{Perm}(U_{in,out}*{U^{\dagger}_{in,out}}_{1,\sigma}),
\end{equation}
with 
\begin{equation}
    S = 
    \begin{pmatrix}
        1 & x \\
        x & 1
    \end{pmatrix}, \ 
    x = \braket{\psi_1|\psi_2},
\end{equation}
where $x$ characterizes the indistinguishability between two photons. Two indistinguishable photons are described by $x=1$ and equation \ref{prob_disting} turns into equation \ref{prob_indist}. Two completely distinguishable photons are descbied with $x=0$.

In our case of two partially indistinguishable photons equation \ref{prob_disting} can be explicitly written as:
\begin{gather*}
    P = P(\ket{\mathrm{In}}, \ket{\mathrm{Out}}, x) = \\
    |a|^2|d|^2 + |b|^2|c|^2 + x^2(ad(bc)^{*} + bc(ad)^{*}),
\end{gather*}
where 
\begin{equation*}
    U_{In,Out} = 
    \begin{pmatrix}
        a & b \\
        c & d
    \end{pmatrix}.
\end{equation*}

Eventually, to simulate the experimentally measured two photon coincidence count rates we estimated four coefficients $C_j$. We used a multinomial distribution with mean values $\braket{C_j}$ to sample outcome configurations:
\begin{equation*}
    \braket{C_j} = \nu T_{exposure} P_j,
\end{equation*}
where $\nu$ is the two photon generation rate, $T_{exposure}$ -- the exposure time, and $P_j$ -- the probability of obtaining the corresponding Fock state calculated according to (\ref{prob_disting}). The sampled set of configurations was used to estimate the $C_{j}$ coefficients.

Process $\chi$-matrix obtained from the simulation with ideal setup parameters is depicted in Fig \ref{fig::gates_fidelity} b).

\begin{figure*}[ht!]
\centering
\includegraphics[width=2.0\columnwidth]{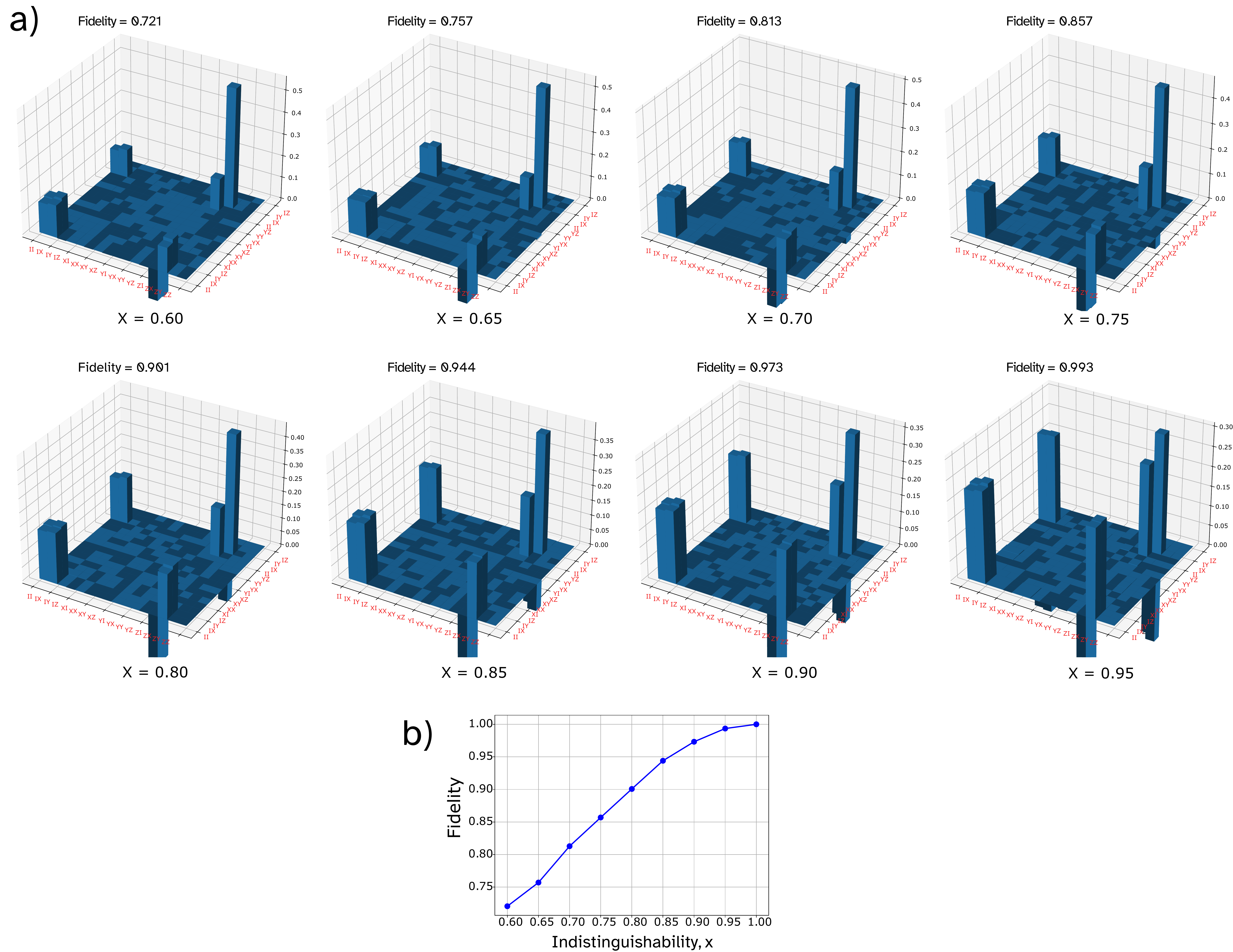}
\caption{\label{fig::sim_indist} Numerical simulation of the influence of different mutual photons indistinguishability on the resulting process $\chi$-matrix obtained from the quantum process tomography procedure and its' fidelity with the ideal two qubit CNOT gate process $\chi$-matrix. Only real part of $\chi$-matrix is shown as imaginary part is negligibly small. The negative values correspond to the elements which phase is close to $pi$.}
\end{figure*}

\begin{figure*}[ht!]
\centering
\includegraphics[width=2.0\columnwidth]{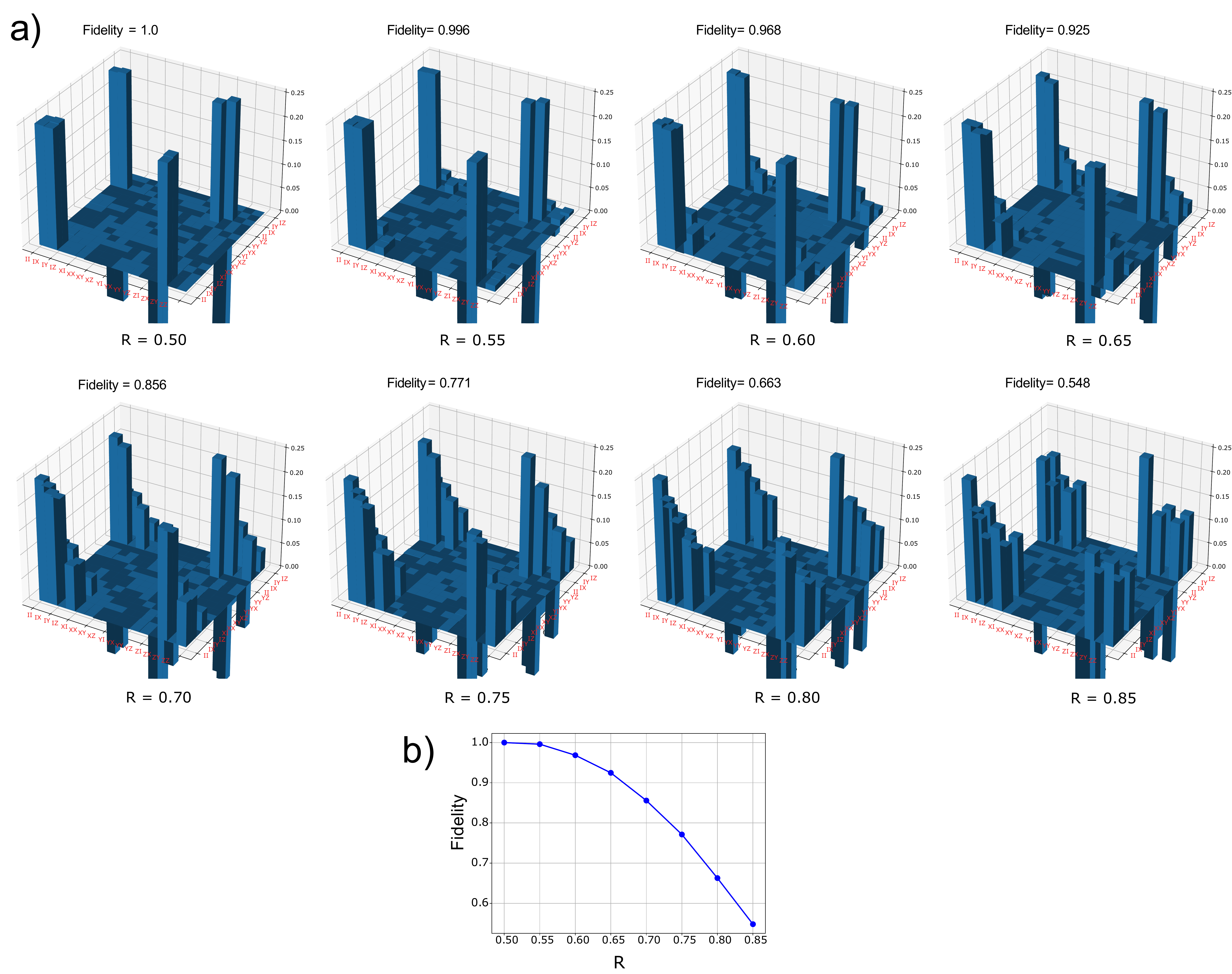}
\caption{\label{fig::sim_R_CNOT_50_59} Numerical simulation of the influence of imperfect splitting ratio of the directional couplers $R_5$ and $R_9$ in the linear optical CNOT gate on the resulting process $\chi$-matrix obtained from the quantum process tomography procedure and its' fidelity with the ideal two qubit CNOT gate process $\chi$-matrix. Only real part of $\chi$-matrix is shown as imaginary part is negligibly small. The negative values correspond to the elements which phase is close to $pi$.}
\end{figure*}

\begin{figure*}[ht!]
\centering
\includegraphics[width=2.0\columnwidth]{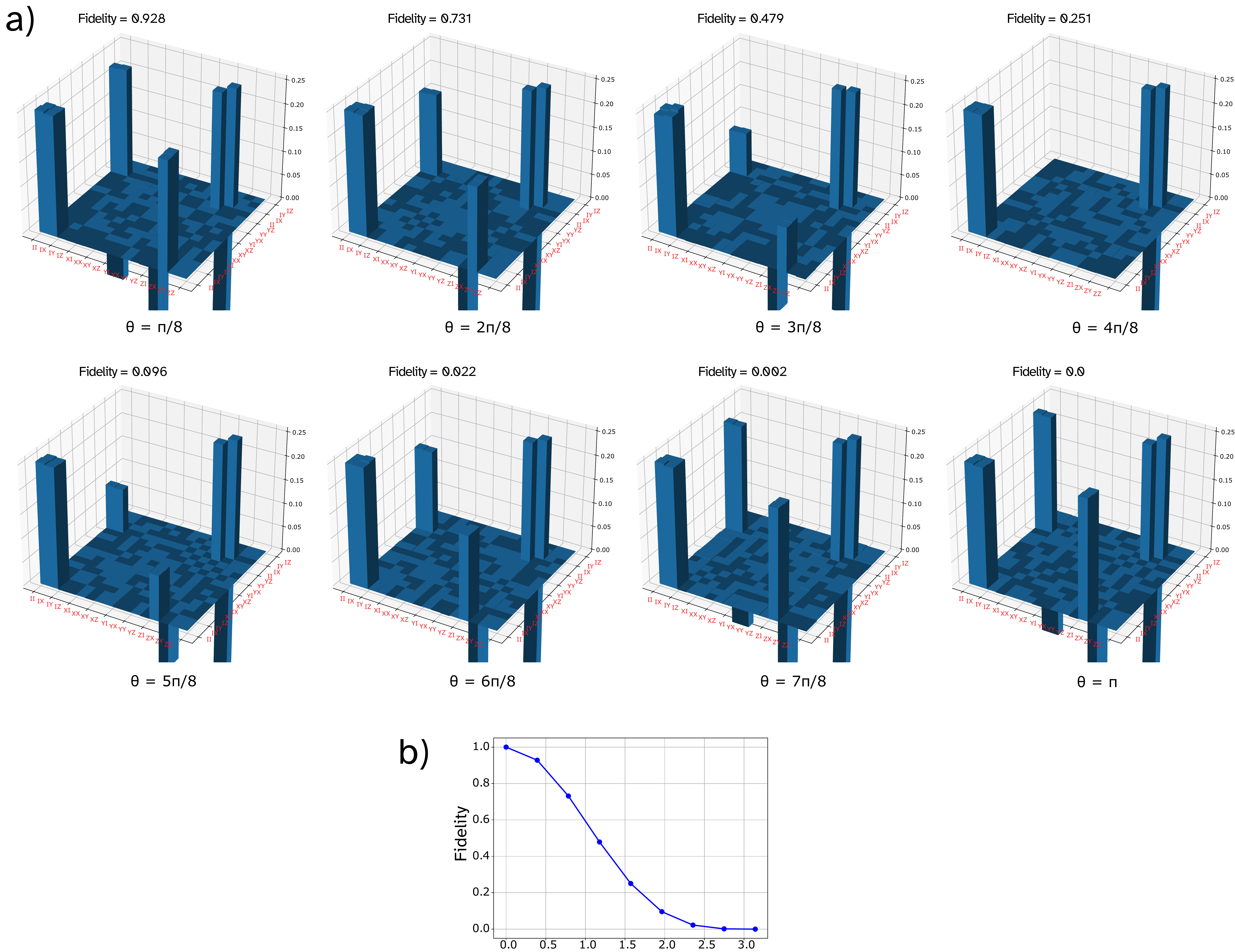}
\caption{\label{fig::sim_theta_CNOT} Numerical simulation of the influence of nonzero $\theta_1$ phase in the linear optical CNOT gate on the resulting process $\chi$-matrix obtained from the quantum process tomography procedure and its' fidelity with the ideal two qubit CNOT gate process $\chi$-matrix. Only real part of $\chi$-matrix is shown as imaginary part is negligibly small. The negative values correspond to the elements which phase is close to $pi$.}
\end{figure*}

\begin{figure*}[ht!]
\centering
\includegraphics[width=2.0\columnwidth]{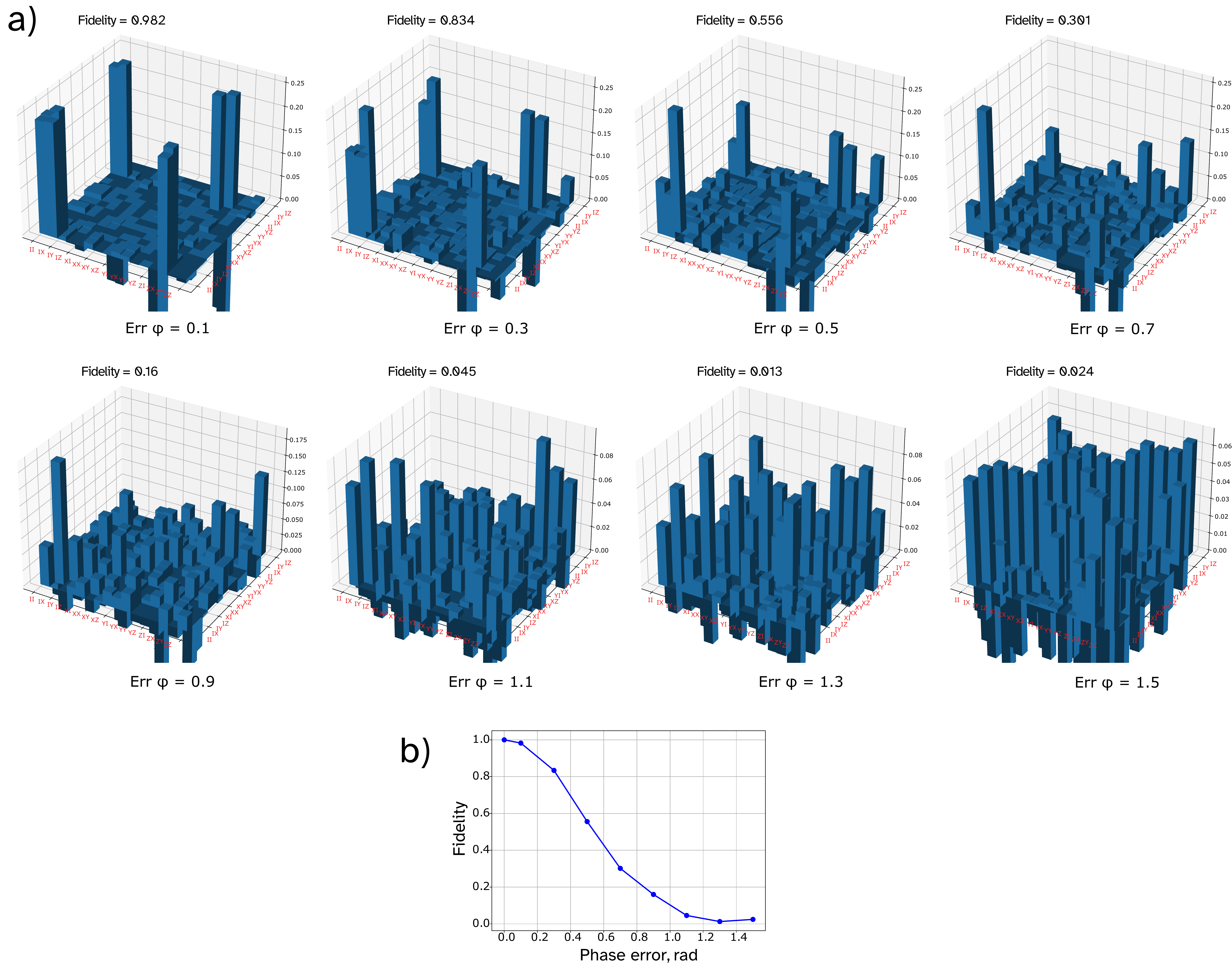}
\caption{\label{fig::sim_phases_set} Numerical simulation of the influence of systematical error in setting phases $\varphi_1$ - $\varphi_8$ in the input and output single qubit gates on the resulting process $\chi$-matrix obtained from the quantum process tomography procedure and its' fidelity with the ideal two qubit CNOT gate process $\chi$-matrix. Only real part of $\chi$-matrix is shown as imaginary part is negligibly small. The negative values correspond to the elements which phase is close to $pi$.}
\end{figure*}

We simulated the experimental data obtained for different setup parameter deviations (power splitting ratios $R_{1-13}$ of the directional couplers, photons indistinguishability $x$, etc.) and used them to extract the $\chi$-matrix similarly to the actual experiment.
We numerically investigated the influence of different types of experimental imperfections on the resulting process $\chi$-matrix obtained from the quantum process tomography procedure and its' fidelity with the ideal two qubit CNOT gate process $\chi$-matrix. Influence of imperfect mutual indistinguishability of two input photons, errors in directional couplers splitting ratios, nonzero $\theta_1$ static phase and systematical errors in single qubit gates phases setting are shown in figures \ref{fig::sim_indist}, \ref{fig::sim_R_CNOT_50_59}, \ref{fig::sim_theta_CNOT} and \ref{fig::sim_phases_set}.

\section{Variational eigenvalue solver}\label{app:vqe}

The variational eigenvalue solver on chip was experimentally implemented using two indistinguishable photons from the SPDC source. The photons were launched into the second and fourth input mode of the optical chip (see Fig.~\ref{fig::chip_structure_act}) initializing two qubit state $\ket{00}$, which was then transformed into probe state $\ket{\Psi}$ by input single-qubit gates $R_{x1}, R_{z1}$ and $R_{x2}, R_{z2}$ and by two-qubit CNOT gate and, lastly, specific projective measurements (defined by the Hamiltonian upon investigation) of $\ket{\Psi}$ were made with the help of output single-qubit gates $R_{x3}, R_{z3}$ and $R_{x4}, R_{z4}$ in order to estimate the Hamiltonian's expectation value $\bra{\Psi}H\ket{\Psi}$. Variational quantum eigenvalue solver algorithm stands for utilizing classical optimization routine to minimize the measured Hamiltonian's expectation value $\bra{\Psi}H\ket{\Psi}$ by varying parameters affecting $\ket{\Psi}$, which are tunable phase shifts $\varphi_1, \varphi_2, \varphi_3, \varphi_4$ in the experiment.

During the algorithm operation, each qubit was measured in one of the bases: $\ket{0}, \ket{1}$ or HV (analogous to polarization qubit encoding notation), $\ket{D},\ket{A}$ (diagonal and anti-diagonal), or $\ket{R},\ket{L}$  (right and left circular) as it was in the quantum process tomography of the CNOT gate (qubits were dual-rail encoded). To make it easier for the algorithm to estimate the proximity of the obtained energy value to the minimum, the input Hamiltonian is immediately represented as a superposition of projections onto the indicated bases. It is convenient that, for example, in the HV basis only the terms with Pauli matrices $Z$ and $I$ will be nonzero; similarly, for DA, the terms with matrix $X$ will remain, and for LR basis -- $Y$, which follows from the Pauli matrices representation through the projection operators:
\begin{gather*}\label{eqn:vqe_pauli_via_projectors}
  \mathbb{I} = P_H + P_V, \ \sigma_X = P_D - P_A , \\
  \sigma_Y = P_R - P_L, \ \sigma_Z = P_H - P_V,
\end{gather*}
where $P_{\psi} = \ket{\psi} \bra{\psi}$ - projection operator onto state $\ket{\psi}$.

Thus, Hamiltonian of $H_2$ molecule written in (\ref{eqn:qubit_hamiltonian}) in the projector operators basis would become:
\begin{eqnarray}\label{eqn:vqe_hamiltonian_projectors}
  H_{H_2} = \tilde{f_0} P_H \otimes P_H + \tilde{f_1} P_H \otimes P_V + \tilde{f_2} P_V \otimes P_H + 
  \nonumber \\ \tilde{f_3} P_V \otimes P_V + \tilde{f_4} P_D \otimes P_D + \tilde{f_5} P_D \otimes P_A + 
  \nonumber \\  \tilde{f_6} P_A \otimes P_D + \tilde{f_7} P_A \otimes P_A ,
\end{eqnarray}
with coefficients $\tilde{f_i}$ calculated from $f_j$ as:
\begin{eqnarray}\label{eqn:vqe_coefficient_relation}
  \tilde{f_0} = f_0 + f_1 + f_2 + f_3, \  \tilde{f_1} = f_0 - f_1 + f_2 - f_3,
  \nonumber \\ \tilde{f_2} = f_0 - f_1 - f_2 + f_3, \  \tilde{f_3} = f_0 + f_1 - f_2 - f_3, 
  \nonumber \\  \tilde{f_4} = f_4, \  \tilde{f_5} = -f_4, \  \tilde{f_6} = -f_4, \  \tilde{f_7} = f_4,
\end{eqnarray}
and values of coefficients $f_j$ were taken from the Openfermion package. 
As an example, for the Hamiltonian describing $H_2$ molecule with distance $0.4$ A between atoms, the coefficients $\tilde{f_i}$ for nonzero projections are presented in Table \ref{tab::hamiltonian_H2_0_4_A}.

In the experiment, similarly to how it was in the quantum process tomography of the CNOT gate,  all four combinations of the projections were measured simultaneously: in hh basis $C_1$ - HH, $C_2$ - HV, $C_3$ - VH, $C_4$ - VV; and in dd basis $C_1$ - DD, $C_2$ - DA, $C_3$ - AD, $C_4$ - AA.
Therefore, for estimation of the Hamiltonian's expectation value $\langle H \rangle_{e}$, one needs to measure two-fold coincidence counts in the above mentioned bases and sum up normalized counts multiplied by a corresponding coefficient from the Table \ref{tab::hamiltonian_H2_0_4_A}:
\begin{eqnarray}\label{eqn:vqe_expectation_value_estimation}
  \langle H \rangle_{e} = \tilde{f_0} c^{hh}_1 + \tilde{f_1} c^{hh}_2 + \tilde{f_2} c^{hh}_3 + \tilde{f_3} c^{hh}_4 + 
  \nonumber \\  + \tilde{f_4} c^{dd}_1 + \tilde{f_5} c^{dd}_2 + \tilde{f_6} c^{dd}_3 + \tilde{f_7} c^{dd}_4,
  \nonumber \\  c^{hh}_i = C^{hh}_i/\sum C^{hh}_i, \ c^{dd}_i = C^{dd}_i/\sum C^{dd}_i,
\end{eqnarray}
where $C^{hh}_i$ and $C^{dd}_i$ are measured two-fold coincidence counts in bases hh and dd correspondingly.

{\setlength{\extrarowheight}{2pt}%
\setlength\tabcolsep{13pt}
\begin{table*}[htbp!]
    \centering
    \begin{tabular}{ |c|c|c|c|c| } 
     \hline
     hh config & HH ($\tilde{f_0}$) & HV ($\tilde{f_1}$) & VH ($\tilde{f_2}$) & VV ($\tilde{f_3}$)  \\ \hline
     value & 1.851 & 0.447 & 0.447 & -0.904 \\
     \hline
     \hline
     dd config & DD ($\tilde{f_4}$) & DA ($\tilde{f_5}$) & AD ($\tilde{f_6}$) & AA ($\tilde{f_7}$)  \\ \hline
     value & 0.165 & -0.165 & -0.165 & 0.165 \\ \hline
    \end{tabular}
    \caption{\label{tab::hamiltonian_H2_0_4_A} Coefficients $\tilde{f_i}$ for nonzero projections for the Hamiltonian (\ref{eqn:vqe_hamiltonian_projectors}) describing $H_2$ molecule with distance $0.4$ A between atoms. }
\end{table*}

Experimental VQE algorithm convergence example and corresponding evolution of the two-fold coincidence counts are presented in Fig. \ref{fig::vqe_plots_example}.

\begin{figure*}[ht!]
\centering
\includegraphics[width=2.0\columnwidth]{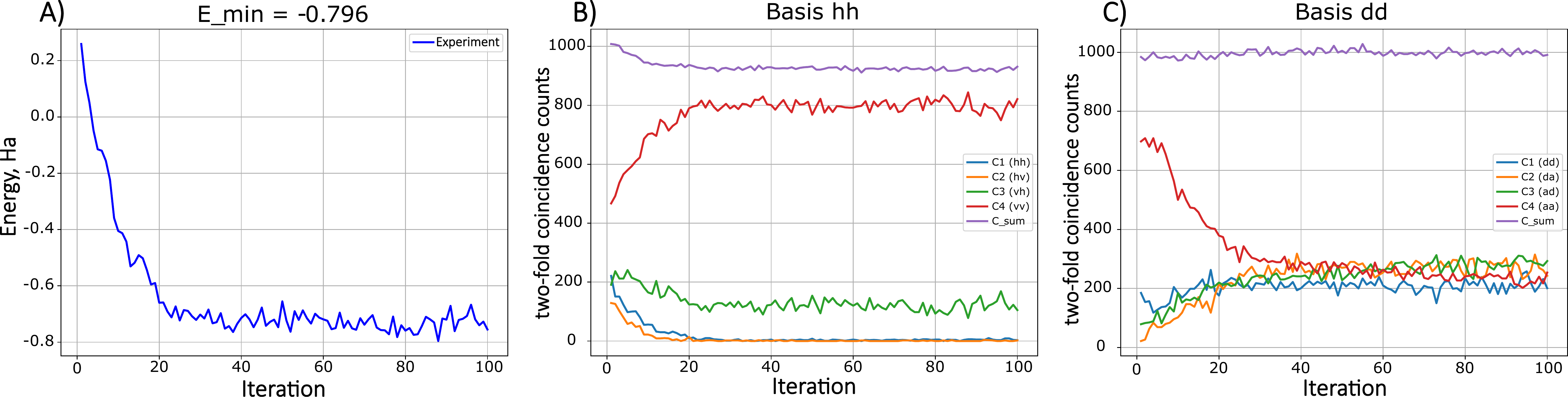}
\caption{\label{fig::vqe_plots_example} Experimental binding energy estimation of a hydrogen molecule $H_2$ with distance $0.4$ A between atoms on a photonic glass chip via the variational quantum eigensolver (VQE) algorithm. A) Experimental VQE algorithm convergence. B) Evolution of two-fold coincidence counts in basis hh during the VQE algorithm. C) Evolution of two-fold coincidence counts in basis dd during the VQE algorithm. }
\end{figure*}

\end{document}